\documentclass[twocolumn,secnumarabic,amssymb, nobibnotes, aps, prd,longbibliography]{revtex4-1}

\usepackage{amsmath}
\usepackage{mathrsfs}
\usepackage{verbatim}
\usepackage{graphicx}
\usepackage{float}
\usepackage{color}

\newcommand\avT[1]{\ensuremath{\langle #1 \rangle_t }}
\newcommand\avIn[1]{\ensuremath{\langle #1\rangle_{\mathrm{init}} }} 
\newcommand\avJ[1]{\ensuremath{[ #1 ]_J }}
\newcommand{\jor}{{\mathrm{jor}}}

\newcommand{\erfc}{{\mathrm{erfc}}}

\begin{document}

\title{How strong are correlations in strongly recurrent neuronal networks?}%

\author{Ran Darshan$^{1,2,3}$, Carl van Vreeswijk$^3$ and David Hansel$^3$}%
\email[Correspondance author:]{david.hansel@univ-paris5.fr}
\affiliation{1. ELSC, The Hebrew University of Jerusalem, Israel;\\ 
2. Janelia Research Campus, Howard Hughes Medical Institute, Ashburn, VA, USA;\\ 
3. Center for Neurophysics, Physiology and Pathology; Cerebral Dynamics, Plasticity and Learning Research Team; CNRS-UMR8119, Paris, France
}
%\date{\today}%
\maketitle
%\tableofcontents

\section*{Abstract}
Cross-correlations in the activity in neural networks are commonly used to characterize their dynamical states and their anatomical and functional organizations. Yet, how these latter network features affect the spatio-temporal structure of the correlations in recurrent networks is not fully understood. Here, we develop a general theory for the emergence of correlated neuronal activity from the dynamics in strongly recurrent networks consisting of several populations of binary neurons. We apply this theory to the case in which the connectivity depends on the anatomical or functional distance between the neurons. We establish the architectural conditions under which the system settles into a dynamical state where correlations are strong, highly robust and spatially modulated. We show that such strong correlations arise if the network exhibits an {\it effective} feedforward structure. We establish how this feedforward structure determines the way correlations scale with the network size and the degree of the connectivity. In networks lacking an effective feedforward structure correlations are extremely small and only weakly depend on the number of connections per neuron. Our work shows how strong correlations can be consistent with highly irregular activity in recurrent networks, two key features of neuronal dynamics in the central nervous system.

\section{Introduction}
Two-point correlations are commonly used to characterize collective dynamics in extended systems \cite{glauber1963time,cross1993pattern,hansel1996chaos,seifert2012stochastic,diakonos2014universal,doiron2016mechanics,tchumatchenko2010correlations}. Recent technical advances \cite{maynard1997utah,peron2015comprehensive,rossant2016spike} make it possible to simultaneously record the activity of many neurons in networks in the brain. This allows for the measurement of two-point correlations for large numbers of neuronal pairs in spontaneous activity as well as upon sensory stimulation or in controlled behavioral conditions. 

Correlations in neuronal activity impact the ability of networks to encode information \cite{zohary1994correlated,abbott1999effect,sompolinsky2001population,averbeck2006neural}. Correlations are also functionally important in performing sensory, motor, or cognitive tasks \cite{buzsaki2006rhythms}. For instance, correlated oscillatory activity has been hypothesized to be involved in visual perception \cite{gray1989oscillatory}. In a recent study combining modeling, electrophysiology and analysis of behavior, we argued that non-oscillatory correlated neuronal activity in the central nervous system is essential in the generation of exploratory behavior \cite{darshan2017canonical}. Correlations are also important for the self-organization  of neuronal networks through activity dependent plasticity \cite{tannenbaum2016shaping}. Indeed, changes in synaptic strength are thought to depend on the temporal correlation of the activty of the pre and postsynaptic neurons \cite{hebb2005organization,bi2001synaptic}.

Correlation strengths depend on the brain area, \cite{ecker2010decorrelated,cohen2011measuring}, the layer in cortex \cite{hansen2012correlated}, stimulus conditions,  behavioral states \cite{dadarlat2017locomotion} and experience \cite{rothschild2013elevated,komiyama2010learning,jeanne2013associative}.  A wide range of values for correlation coefficients, from negligible \cite{ecker2010decorrelated,renart2010asynchronous} to substantial \cite{gawne1993independent,gawne1996adjacent,smith2008spatial,zohary1994correlated,gutnisky2008adaptive,bair2001correlated} have been reported in the last two decades. Correlation coefficients are usually higher for close-by neurons than for neurons that are far apart \cite{lee1998variability,smith2008spatial,levy2012spatial,fino2011dense,tanaka2014spatial}. In cortex, they drop significantly over distances of $200-400$~$\mu m$ \cite{levy2012spatial,tanaka2014spatial}. Recent works have reported correlations varying non-monotonically with distance \cite{safavi2017non} or correlations which are positive for close-by neurons but negative for neurons farther apart \cite{rosenbaum2017spatial}. Correlation coefficients also depend on functional properties of the neurons. Neurons which code for similar features of sensory stimuli are more correlated \cite{smith2008spatial,lee1998variability,denman2014structure,ecker2010decorrelated,tanaka2014spatial}. 

Neurons in cortex receive recurrent inputs from several hundreds \cite{binzegger2004quantitative} to a few thousands \cite{abeles1991corticonics} of other neurons. Individual connections can induce post-synaptic potentials in a range of 0.1~mV to several mV \cite{levy2012spatial}. Thus, $50$ simultaneous inputs are sufficient to trigger or suppress a spike in a neuron. These facts have been incorporated in model networks with strongly recurrent connectivity in which connection strengths are $\mathcal{O}(1/\sqrt{K})$, where $K$ is the average number of inputs per neuron. This scaling is in contrast to the one used in standard mean-field models (e.g. \cite{ginzburg1994theory}) where connections  are $\mathcal{O}(1/K)$ and thus are much weaker. Recent experiments in cortical cultures are consistent with a $1/\sqrt{K}$ scaling of connection strength  \cite{barral2016synaptic}.

van Vreeswijk and Sompolinsky \cite{van1996chaos,vreeswijk1998chaotic} showed that strongly recurrent networks consisting of two populations of neurons, one excitatory (E) and one inhibitory (I), randomly connected on a directed Erd\"{o}s-R\'{e}nyi graph, operate in a state in which the strong excitation is balanced by the strong inhibition. In this {\it balanced regime}, neurons receive strong excitatory and inhibitory inputs, each $\mathcal{O}(\sqrt{K})$, but due to the recurrent dynamics these inputs cancel each other at the leading order. This cancellation, which does not require fine-tuning, results in $\mathcal{O}(1)$ net inputs into the neurons whereas spatial heterogeneities and temporal fluctuations in the inputs are also $\mathcal{O}(1)$. As a result, the activity of the neurons remains finite and exhibits strong temporal irregularity and heterogeneity \cite{monteforte2012dynamic,roxin2011distribution}. 

The latter results were established in two-population sparsely connected networks. Renart et al. \cite{renart2010asynchronous} showed that they also hold for densely connected networks. This is because in unstructured and strongly recurrent networks the dynamics suppress correlations \cite{helias2014correlation}. Despite the fact that in these networks neurons share a finite fraction of their inputs, they operate in an asynchronous state with  $\mathcal{O}(1/N)$ correlations, where $N$ is the number of neurons in the network \cite{hansel1992synchronization}. 

These previous studies focused on strongly recurrent unstructured networks with two neuronal populations. In the brain, however, neural networks comprise  a diversity of excitatory and inhibitory cell types, which differ in their morphology, molecular signature and importantly for our purpose, in their connectivity. These networks are also structured at many levels. In particular, the probability of connection  falls off with distance and depends on functional properties of pre- and postsynaptic neurons. For example, in mouse primary auditory cortex the probability of excitatory neurons to be connected decays to zero within $\sim 300 \mu m$ \cite{levy2012spatial}. In cat primary visual cortex neurons interact locally on range of  $\sim 500 \mu m$, whereas long range patchy connections are observed up to several $mms$ \cite{kisvarday1997orientation,stepanyants2009fractions}. 

§§In the present paper we investigate how {\it structure} in network connectivity can give rise to strongly correlated activity. Our goal is to explore the general architectural features that control the strength of pair-wise correlations in strongly recurrent neural circuits consisting of several excitatory and inhibitory neuronal populations. In Section~\ref{SModel}, we define the network architectures and the neuronal model we use. In Section~\ref{SSpatiotemporal}, we establish a set of constraints, {\it the balanced correlation equations}, that need to be satisfied in any strongly recurrent network if firing rates do not saturate. We derive in Section~\ref{S2networks} explicit expressions for the Fourier components of the correlations in two-population networks with spatially modulated connectivity. We establish the conditions under which correlations are strong and show that these do not violate the balanced correlation equations. Section~\ref{Stheorem} is devoted to networks with an arbitrary number of populations. We prove two theorems, which state for which network architectures correlations are $\mathcal{O}(1/N)$ and for which they increase with $K$, when $K$ is large. In Section~\ref{AppliTh}, we apply these theorems to specific examples. In Section~\ref{SlargeK}, we assume that $K=\mathcal{O}(N^{\gamma})$ and derive a bound on $\gamma$ for which the scaling theorems still apply. The paper closes with a discussion of our results.
 
\section{The network model}\label{SModel}

\subsection{Architecture}
\begin{figure}[htp]
\begin{center}
\includegraphics[clip,width=0.5\textwidth]{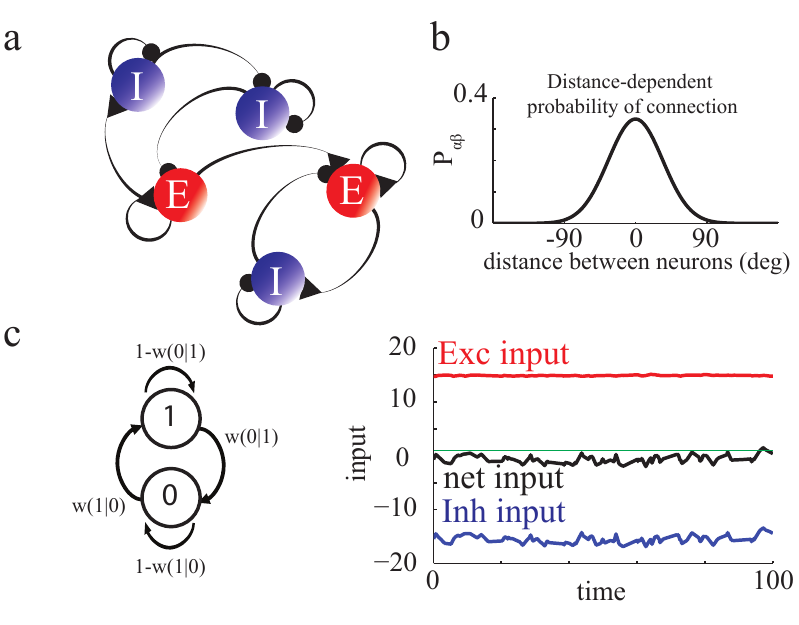}
\end{center}
\caption{{\bf Network architectures and dynamics.} {\bf a.} Networks consist of $D$ populations of neurons, excitatory ($E$,red) and inhibitory ($I$, blue), recurrently coupled and receiving external drives. Triangles: Excitatory connections. Circles: Inhibitory connections. {\bf b.} The neurons in each population are located on a ring. The probability that 
neuron $(j,\beta)$ is connected to neuron $(i,\alpha)$ depends on their distance. {\bf c.} Left: Neurons are binary units with zero temperature Glauber dynamics.
Right: The network operates in the balanced regime. Red: Excitatory input into a neuron. Blue: Inhibitory input into the same neuron. Black: The net input is on the order of the threshold (green, $T=1$). Time is given in units of $\tau=1$.}
\label{fig:Model}
\end{figure}

We consider a neuronal network with a ring architecture, comprising $D$ neuronal populations, some excitatory and others inhibitory 
(Fig.~\ref{fig:Model}; \cite{ben1997traveling,hansel199813}). 
For simplicity, we assume that all populations have the same number of neurons, $N$. 
Neuron $i$, in population $\alpha$ (neuron (i,$\alpha$)) is located at angle $\theta_i^\alpha=\frac{2\pi i}{N}$ with, $i=1,\ldots,N$ and $\alpha=1,\ldots,D$.  
The probability, $P_{ij}^{\alpha\beta}$, that 
a neuron $(j,\beta)$  projects to neuron $(i,\alpha)$ depends on their 
distance on the ring (Fig.~\ref{fig:Model}b). 
We write: $P_{ij}^{\alpha\beta}= \frac{K}{N}f_{\alpha\beta}(|\theta_i^\alpha-\theta_j^\beta|)$, 
where
$\sum_j f_{\alpha\beta}(|\theta_i^\alpha-\theta_j^\beta|)=N$. In this paper we assume a finite number of non-zero Fourier modes in $f_{\alpha\beta}$.

Thus, a neuron in population $\alpha$ receives, on average, $K$ 
inputs from neurons in each of the populations $\beta$\@. 
We denote by $\boldsymbol{\Lambda}$ the adjacency matrix of the network connectivity
\[
	\Lambda_{ij}^{\alpha\beta}=\Bigg\{\begin{array}{ll}
       1 & \mbox{if ($j,\beta$) is presynaptic to ($i,\alpha$)} \\
       0 &  		\mbox{otherwise}				\end{array} 
        \]
For simplicity we assume that all connections from population $\beta$ to population $\alpha$ have the same strength, $j_{\alpha\beta}$\@. 
We thus define the connectivity matrix, $\boldsymbol{J}$, as
\[	J^{\alpha\beta}_{ij}=j_{\alpha\beta}\Lambda_{ij}^{\alpha\beta}.
\]
Note that  $j_{\alpha\beta}$ is positive (negative) if population $\beta$ is excitatory (inhibitory).

In all this paper we focus on strongly recurrent networks, characterized by interactions which are $\mathcal{O}(\frac{1}{\sqrt{K}})$ \cite{van1996chaos,vreeswijk1998chaotic}. Thus, we scale the synaptic strengths with the mean connectivity as  
\begin{equation}
j_{\alpha\beta}=\frac{J_{\alpha\beta}}{\sqrt{K}}.
\end{equation}
where $J_{\alpha\beta}$ is $\mathcal{O}(1)$.

\subsection{Single neuron dynamics}

The state of neuron $(i,\alpha)$ is characterized by a binary variable, $S_i^\alpha$\@. 
When the neuron is quiescent, $S_i^\alpha=0$, while if it 
is active, $S_i^\alpha=1$. The total synaptic current into neuron $(i,\alpha)$ at time $t$,  $h_i^\alpha(t)$, is 
the result of  all its recurrent interactions 
with the other neurons in the network, as well as the feedforward inputs coming from outside the network. It is given by 
\begin{align}\label{model}
h_i^\alpha(t)=\sum_{\beta=1}^D \sum_{j=1}^N J_{ij}^{\alpha\beta}S_j^\beta(t)+I_i^\alpha
\end{align}
where the external input $I_i^\alpha=\sqrt{K}I_\alpha(\theta)$ \cite{vreeswijk1998chaotic} is 
assumed to be constant in time. 
Here, $I_\alpha$ is $\mathcal{O}(1)$ and thus $I_i^\alpha$ is $\mathcal{O}(\sqrt{K})$. 

The neurons are updated with Glauber dynamics at zero temperature \cite{glauber1963time,ginzburg1994theory,renart2010asynchronous}. 
Specifically, update times for neuron $(i,\alpha)$ are Poisson distributed with rate $1/\tau$: if neuron $(i,\alpha)$ is updated at time $t$, $S_i^\alpha$ is set to $1$ if $h_i^\alpha(t) \geq T_{\alpha}$ and to $0$ otherwise (for simplicity, we assume the same threshold, $T_{\alpha}$, for all neurons in population $\alpha$,
and the same update rate for all neurons).
Accordingly, the transition probability, $w$, (Fig.~\ref{fig:Model}c) can be written as
\[
	w(S_i^\alpha(t)\to 1-S_i^\alpha(t))=\frac{1}{\tau}[S_i^\alpha(t)-\Theta(h_i^\alpha(t)-T_{\alpha})]^2.
\]
where $\Theta$ is the Heaviside function. We normalize time so that $\tau=1$.

\section{Spatio-temporal profile and correlations of the activity for $\boldsymbol{N,K \to \infty}$} \label{SSpatiotemporal}

We write the state of neuron ($i,\alpha$) as
\begin{equation}\label{Sseparate}
S_i^\alpha(t)=m_\alpha(\theta_i^\alpha)+\Delta S_i^\alpha+
\delta S_i^\alpha(t),
\end{equation} where $m_{\alpha}(\theta)=\avJ{\avT{S_i^\alpha(t)}}$ is the average of $S_i^\alpha(t)$ over realizations of the network connectivity matrix and over time. In Eq.~(\ref{Sseparate}), the term $\Delta S_i^\alpha$  is the quenched disorder, $\Delta S_i^\alpha=\avT{S_i^\alpha(t)}-m_\alpha(\theta_i^\alpha)$, whereas $\delta S_i^\alpha(t)$ represents the temporal fluctuations in the activity of neuron $(i,\alpha)$, $\delta S_i^\alpha(t)=S_i^\alpha(t)-\avT{S_i^\alpha(t)}$.

Similarly, we can write:
\begin{equation}\label{decomph}
h_i^\alpha(t)=h_\alpha(\theta_i^\alpha)+\Delta h_i^\alpha
        +\delta h_i^\alpha(t),
\end{equation}
where $h_\alpha(\theta_{i}^{\alpha}) =\avJ{\avT{h_{i}^{\alpha}}}$ is a smooth function of its argument. For large $N$ it is given by:
\begin{equation}\label{eqquenchh}
h_\alpha(\theta)=\sqrt{K}\left[
\sum_{\beta=1}^{D}J_{\alpha\beta}\int 
\frac{d\theta^\prime}{2\pi}f_{\alpha\beta}(\theta-\theta')m_{\beta}(\theta^\prime)+I_\alpha(\theta)\right].
\end{equation}

The second term in Eq.~\eqref{decomph} is the quenched disorder in the input. It satisfies
\begin{equation}\label{eqDeltah}
\Delta h_i^\alpha=\sum_{\beta=1}^{D}\frac{J_{\alpha\beta}}{\sqrt{K}}\sum_j \left[\Delta \Lambda_{ij}^{\alpha\beta}m_\beta(\theta_j^\beta)+
\Lambda_{ij}^{\alpha\beta}\Delta S_j^\beta\right],
\end{equation}
where $\Delta\Lambda_{ij}^{\alpha\beta}=\Lambda_{ij}^{\alpha\beta}-\avJ{\Lambda_{ij}^{\alpha\beta}}$.

Finally, the temporal fluctuations in the inputs are 
\begin{equation}\label{deltah}
\delta h_i^\alpha(t)=
\sum_{\beta=1}^D\frac{J_{\alpha\beta}}{\sqrt{K}}\sum_j
        \Lambda_{ij}^{\alpha\beta}\delta S_j^\beta(t).
\end{equation}
Because we scale the synaptic strength as $1/\sqrt{K}$, these temporal fluctuations
are $\mathcal{O}(1)$ when correlations are weak. At first sight, $h_{\alpha}(\theta)$ is $\mathcal{O}(\sqrt{K})$. This would imply that, depending on whether $h_{\alpha}(\theta)$ is positive or negative, the neurons fire either at very high or at very low rate. This happens unless the network settles into a state in which excitation and inhibition are, on average, balanced. In this case, the net average inputs to the neurons are in fact $\mathcal{O}(1)$. 
On the other hand, large spatial and temporal correlations  could in principle lead to temporal fluctuations and heterogeneities in the inputs of $\mathcal{O}(\sqrt{K})$. Thus in presence of strong correlations it is not sufficient to require that the mean input, $h_{\alpha}(\theta)$, is  $\mathcal{O}(1)$, for the network to operate in a biologically relevant regime. For that, we need  both the mean net inputs and the input fluctuations to be $\mathcal{O}(1)$ at any time and for all neurons. When this happens we say that the network operates in the balanced regime.

\subsection{Balance equations for the quenched averaged population activities}

As in \cite{van2005irregular,rosenbaum2014balanced}, the requirement that the mean input is  $\mathcal{O}(1)$ yields 
\begin{equation}\label{instantbal}
\sum_{\beta=1}^D J_{\alpha\beta}\int\frac{d\theta^\prime}{2\pi}\,f_{\alpha\beta}(\theta-\theta^\prime)m_{\beta}(\theta^\prime)+I_{\alpha}(\theta) = \mathcal{O}(1/\sqrt{K}),
\end{equation}
for all $\alpha\in \{1,\ldots,D\}$ and all $\theta\in[0,2\pi)$.
In the large $K$ limit this yields a set of linear equations which determines the functions $m_{\alpha}(\theta)$ to leading order. These equations can be written in Fourier space 
\begin{equation}\label{BalanceMode}
\sum_{\beta}J^{(n)}_{\alpha\beta}m^{(n)}_{\beta}+I^{(n)}_{\alpha}=\mathcal{O}\left(\frac{1}{\sqrt{K}}\right).
\end{equation}
where the superscript $n$ denotes the $n^{th}$ Fourier mode and we have used the short hand notation
\begin{equation}
J^{(n)}_{\alpha\beta}= J_{\alpha\beta}f^{(n)}_{\alpha\beta}.
\end{equation}
which is the $n^{th}$ Fourier mode of the connectivity matrix. In what follows, we consider $J^{(n)}_{\alpha\beta}$ as the elements of a $D \times D$ matrix, $\boldsymbol{J}^{(n)}$\@.

The spatial average of the activities, $m_{\alpha}^{(0)}$, must be non-negative for the balanced state to exist. This implies that the parameters $J_{\alpha\beta}$ and the external inputs, $I^{(0)}_\alpha$, must satisfy a set of inequalities. For example, for networks with two populations, one excitatory ($\alpha=E$) and one inhibitory ($\alpha=I$), these inequalities are \cite{vreeswijk1998chaotic}
\begin{equation}\label{eBalanceConditions}
\frac{I^{(0)}_{E}}{I^{(0)}_{I}}>\frac{J_{EI}}{J_{II}}>\frac{J_{EE}}{J_{IE}}.
\end{equation}
In general, Eq.~\eqref{BalanceMode}, also implies additional inequalities that must be satisfied by $f^{(n)}_{\alpha\beta}$ to guarantee that 
$m_{\alpha}(\theta)\geq 0$ for all $\alpha$ and $\theta$ \cite{van2005irregular,rosenbaum2014balanced}. 
In the present work we focus on the case where the external inputs are spatially homogeneous. Thus, $m_{\alpha}^{(n)}=0$ for $n \geq 1$, and the only condition which is required for the balanced state to exist is: $m_{\alpha}^{(0)}\geq 0$. 

To study the stability of the homogeneous balanced state it is useful to introduce the {\it interaction} matrix
\begin{equation}\label{StabMat}
\bar{\mathcal{J}}_{ij}^{\alpha\beta}=g_{\alpha}J_{\alpha\beta}f_{\alpha\beta}(\theta_i^{\alpha}-\theta_j^{\beta})
\end{equation}
where $g_{\alpha}$ is the population averaged gain (see Appendixes \ref{ApCorBinary},\ref{ApSCEqns}; \cite{vreeswijk1998chaotic,helias2014correlation}). 

Small perturbations, $\delta m_{\alpha}(\theta,t)$,  of the activity profile around this homogeneous state evolve according to \cite{vreeswijk1998chaotic}: 
\begin{equation}\label{StabilityPop}
\frac{d\delta m^{(n)}_{\alpha}}{dt}=-\delta m^{(n)}_{\alpha}+\sqrt{K}\sum_{\beta}\bar{\mathcal{J}}^{(n)}_{\alpha\beta}\delta m^{(n)}_{\beta},
\end{equation}
where $\delta m_{\alpha}^{(n)}(t)$ and $\bar{\mathcal{J}}^{(n)}_{\alpha\beta}$ are the $n^{th}$ Fourier modes of  $\delta m_{\alpha}(\theta,t)$ and $\bar{\mathcal{J}}_{ij}^{\alpha\beta}$. Since each row of $\bar{\mathcal{J}}_{ij}^{\alpha\beta}$ has $\mathcal{O}(K)$ non-zero elements which are  $\mathcal{O}(1/K)$, $\bar{\mathcal{J}}^{(n)}_{\alpha\beta}$ is $\mathcal{O}(1)$.

The stability of the balanced state with respect to perturbations in $m_{\alpha}(\theta)$ requires that, for all $n$, all the eigenvalues of the matrices $\boldsymbol{\mathcal{J}}^{(n)}$ have real parts smaller than $1$. 
For instance, for a two population (E,I) network one must have in the large $K$ limit
\begin{equation}
J_{EE}|J_{II}|f^{(n)}_{EE} f^{(n)}_{II}-J_{IE}|J_{EI}|f^{(n)}_{EI}f^{(n)}_{IE} \leq 0.
\end{equation}
for all $n$. Note that the gain of the neurons, $g_{\alpha}$, dropped from this equation. 
The balanced state undergoes a Turing bifurcation when for some $n \geq 1$, $J_{EE}|J_{II}|f^{(n)}_{EE} f^{(n)}_{II}-J_{IE}|J_{EI}|f^{(n)}_{EI}f^{(n)}_{IE}$ crosses 0 and becomes positive.

\subsection{Balance equation for pair-wise correlations}

The time-lagged auto- and cross-correlation functions of the activities of a pair
of neurons, $(i,\alpha), (j,\beta)$, for
$(j,\beta)\neq(i,\alpha)$, are 
\begin{eqnarray}\label{CorDef}
	a^{\alpha}_{i}(\tau)&=&\avT{\delta S_i^\alpha(t)\delta S_i^\alpha(t+\tau)} \\
	c^{\alpha\beta}_{ij}(\tau)&=&\avT{\delta S_i^\alpha(t)\delta S_j^\beta(t+\tau)}
\end{eqnarray}
In what follows, it is convenient to define $c_{ii}^{\alpha\alpha}=0$.

Due to the randomness of the connectivity, the number of excitatory and inhibitory inputs varies from  neuron to neuron, resulting in heterogeneous firing rates between neurons even when they belong to the same population \cite{vreeswijk1998chaotic,roxin2011distribution}. This structural randomness  also results in heterogeneity of the auto-correlations of single neuron activities. It also contributes to the heterogeneity in pair cross-correlations. The latter is further enhanced by the spatial variability in the number of inputs shared by pair of neurons. A full characterization of the distributions of the auto- and cross-correlations is beyond the scope of this paper. Instead, here we will focus on their population averages. 

The population average auto-correlation,
$A_\alpha$, is given by $A_{\alpha}(\tau)=\frac{1}{N}\sum_i a_i^\alpha(\tau)$,
which in the thermodynamic limit is also $A_{\alpha}(\tau) =\avJ{a_i^\alpha(\tau)}$.
With the architecture we use, the probability that neurons $(i,\alpha)$ and $(j,\beta)$ share common inputs from a third neuron, $(k,\gamma)$ is
\begin{eqnarray}\label{common}
	\Pr(\Lambda_{ik}^{\alpha\gamma}=1 \land\Lambda_{jk}^{\beta\gamma}=1)=\nonumber \\
\frac{K^2}{N^2}f_{\alpha\gamma}(\theta_{i}^{\alpha}-\theta_{k}^{\gamma})f_{\beta\gamma}(\theta_{j}^{\beta}-\theta_{k}^{\gamma})
\end{eqnarray}

As is evident from this equation, the number of shared inputs averaged over all pairs of neurons separated by the same distance on the ring, $\Delta$, depends only on $\Delta$. Thus we define the average cross-correlations as $C_{\alpha\beta}(\Delta,\tau)=\langle c_{ij}^{\alpha\beta}(\tau)\rangle$, where  $\langle . \rangle$ denotes the average over pairs with $|\theta_i^\alpha-\theta_j^\beta| = \Delta $. In the thermodynamic limit, this quantity does not depend on the specific realization of the network connectivity matrix.
The Fourier expansion of this function is
\begin{equation}\label{CG-Cor-Fourier}
C_{\alpha\beta}(\Delta,\tau)=\sum^{N-1}_{n=0}C^{(n)}_{\alpha\beta}(\tau)e^{-i n \Delta},
\end{equation}
where $C^{(n)}_{\alpha\beta}(\tau)=\frac{1}{N^2}
\sum_{kj} c_{kj}^{\alpha\beta}(\tau) e^{i n (\theta_k^\alpha-\theta_j^\beta)}$\@.

Equation \eqref{BalanceMode}, which determines the population averaged firing rates, stems from the constraint that the net input in every neuron must be $\mathcal{O}(1)$ when $K$ is large. The condition that for any pair of neurons the correlation in their inputs is finite in that limit leads to another constraint, which we now derive.  

Let us consider the $N D \times N D$ matrix $\boldsymbol{Q}$ defined by
\begin{equation}\label{CorInputsDef}
Q_{ij}^{\alpha\beta}=\avJ{\avT{\delta h_i^\alpha(t) \delta h_j^\beta(t)}},
\end{equation}
for  $(i,\alpha)\neq(j,\beta)$ and $Q^{\alpha\alpha}_{ii}=0$.
Using Eq.\eqref{deltah}, one finds
\begin{equation}\label{CorInputsH}
Q_{ij}^{\alpha\beta}=\sum_{\gamma\gamma^\prime}\frac{J_{\alpha\gamma}J_{\beta\gamma'}}{K}
\sum_{kk^\prime} \avJ{\Lambda_{ik}^{\alpha\gamma}\Lambda_{jk^\prime}^{\beta\gamma^\prime}}\avJ{\avT{\delta S_k^\gamma(t)\delta S_{k^\prime}^{\gamma^\prime}(t)}}
\end{equation} for $(i,\alpha)\neq(j,\beta)$. 
Similarly to $C_{\alpha\beta}(\Delta)$, $Q_{ij}^{\alpha\beta}$ is a function of $\Delta=|\theta_i^{\alpha}-\theta_j^{\beta}|$ only. 

Expanding Eq.\eqref{CorInputsH} in Fourier $(Q_{\alpha\beta}^{(n)}\equiv
\frac{1}{N^2}\sum_{kj}Q_{kj}^{\alpha\beta}e^{in(\theta_k^\alpha-\theta_j^\beta)})$, yields
\begin{align*}
Q_{\alpha\beta}^{(n)}=K\sum_{\gamma\gamma'}J^{(n)}_{\alpha\gamma} J^{(-n)}_{\beta\gamma'}C_{\gamma\gamma'}^{(n)}+\sum_{\gamma}\frac{K}{N}J_{\alpha\gamma}^{(n)}J_{\beta\gamma}^{(-n)} A_{\gamma}
\end{align*}
which can be written in matrix form
\begin{equation}\label{CorInputs}
\boldsymbol{Q}^{(n)}=K\boldsymbol{J}^{(n)}\boldsymbol{C}^{(n)}[\boldsymbol{J}^{(n)}]^{\dagger}+\frac{K}{N}\boldsymbol{J}^{(n)}\boldsymbol{A}[\boldsymbol{J}^{(n)}]^{\dagger}
\end{equation}
Here $\boldsymbol{X}$ denotes the $D\times D$ matrix $X_{\alpha\beta}$, 
$\boldsymbol{A}_{\alpha\beta}=A_{\alpha}\delta_{\alpha\beta}$ and the superscript $\dagger$ denotes Hermitian conjugation ($[\boldsymbol{J}_{\alpha\beta}^{(n)}]^{\dagger}=[\boldsymbol{J}_{\beta\alpha}^{(-n)}]$).

Note that the diagonal of the matrix $\boldsymbol{Q}$ is not the variance of the inputs. The latter is (see Appendix E4)
\begin{align}
\sigma_\alpha^2=K\sum_{n\gamma\gamma'}J^{(n)}_{\alpha\gamma} J^{(n)}_{\alpha\gamma'}C_{\gamma\gamma'}^{(n)}+\sum_{\gamma}J_{\alpha\gamma}^2 A_{\gamma} 
\end{align}

The requirement that crosscorrelations of the inputs into pair of neurons are at most $\mathcal{O}(1)$ implies that all the quantities $Q_{\alpha\beta}^{(n)}$ are also at most $\mathcal{O}(1)$. This yields
\begin{equation}\label{BalanceCorr}
\boldsymbol{J}^{(n)}\boldsymbol{C}^{(n)}[\boldsymbol{J}^{(n)}]^{\dagger}=\mathcal{O}(1/K)
\end{equation}

We call Eq.~\eqref{BalanceCorr}, the {\it balanced correlation equation}.  
It implies that for all $n$ for which the matrix $\boldsymbol{J}^{(n)}$ is invertible and has entries $\mathcal{O}(1)$, $\boldsymbol{C}^{(n)}$ is smaller or equal to $\mathcal{O}(1/K)$ and is thus weak. For a broad class of network architectures, we show below that correlations are in fact $\mathcal{O}(1/N)$ and barely depend on $K$, for large $K$. When, however, $\boldsymbol{J}^{(n)}$ is singular for some $n\geq 0$, correlations can be larger than $\mathcal{O}(1/K)$. In fact we will see that in those cases correlations can even increase with $K$.

Finally, we note that one can also write a balanced correlation equation for the quenched disorder  of the neural activity, using the fact that $\avJ{\Delta h_i^\alpha\Delta h_j^\beta}^{(n)}$ also must be at most $\mathcal{O}(1)$. This requirement leads to the condition
\begin{equation}\label{BalanceCorrQuench}
\boldsymbol{J}^{(n)}\boldsymbol{\Gamma}^{(n)}[\boldsymbol{J}^{(n)}]^{\dagger}=\mathcal{O}(1/K)
\end{equation}
with
\begin{equation}\label{QuenchDef}
[\boldsymbol{\Gamma}^{(n)}]_{\alpha\beta}\equiv\avJ{\Delta S_i^\alpha\Delta S_j^\beta}^{(n)}
\end{equation}

\section{Correlations in two-population networks}\label{S2networks}

In Appendix~\ref{ApCorBinary} we derive an equation for the spatial Fourier modes of the equal-time quenched average correlation functions, $C_{\alpha\beta}(\Delta,0)$\@. It yields (omitting the second argument)
\begin{eqnarray}\label{CorrEqMat}
2\boldsymbol{C}^{(n)} =\sqrt{K}\left(\boldsymbol{\bar{\mathcal{J}}}^{(n)}\boldsymbol{C}^{(n)}+\boldsymbol{C}^{(n)}[\boldsymbol{\bar{\mathcal{J}}}^{(n)}]^{\dagger}\right)
\\\nonumber
+\frac{\sqrt{K}}{N}\left(\boldsymbol{\bar{\mathcal{J}}}^{(n)}\boldsymbol{A}+\boldsymbol{A}[\boldsymbol{\bar{\mathcal{J}}}^{(n)}]^{\dagger}\right),
\end{eqnarray} 
There we also show that the solution of this equation is a fixed point of the dynamics of the correlations which is always stable when the balanced state is stable with respect to perturbation in the population rates.

This equation holds for a network with an arbitrary number of neuronal populations. In the case of two populations, it can be solved explicitly, yielding after a straightforward calculation (see Appendix \ref{Ap2pop})

\begin{widetext}
\begin{eqnarray}\label{Corfull}
\nonumber
C^{(n)}_{EE}&=&-\frac{A_E}{N}+\frac{1}{N}\frac{-2A_E+A_E\left(T^{(n)}+2\bar{\mathcal{J}}_{II}^{(n)}\right)\sqrt{K}-\left(A_E(\Delta^{(n)}+(\bar{\mathcal{J}}_{II}^{(n)})^2)+A_I(\bar{\mathcal{J}}_{EI}^{(n)})^2\right)K}{-2+3T^{(n)}\sqrt{K}-\left((T^{(n)})^2+2\Delta^{(n)}\right)K+T^{(n)}\Delta^{(n)} K^{3/2}}
\\ \nonumber
C^{(n)}_{EI}&=&-\frac{1}{N}\frac{\left(A_E\bar{\mathcal{J}}_{IE}^{(n)}+A_I\bar{\mathcal{J}}_{EI}^{(n)}\right)\sqrt{K}-\left(A_I\bar{\mathcal{J}}_{EE}^{(n)}\bar{\mathcal{J}}_{EI}^{(n)}+A_E\bar{\mathcal{J}}_{II}^{(n)}\bar{\mathcal{J}}_{IE}^{(n)}\right)K}{-2+3T^{(n)}\sqrt{K}-\left((T^{(n)})^2+2\Delta^{(n)}\right)K+T^{(n)}\Delta^{(n)} K^{3/2}}
\\ \nonumber
C^{(n)}_{II}&=&-\frac{A_I}{N}+\frac{1}{N}\frac{-2A_I+\left(A_I(T^{(n)}+2\bar{\mathcal{J}}_{EE}^{(n)}\right)\sqrt{K}-\left(A_I(\Delta^{(n)}+(\bar{\mathcal{J}}_{EE}^{(n)})^2)) +A_E(\bar{\mathcal{J}}_{IE}^{(n)})^2\right)K}{-2+3T^{(n)}\sqrt{K}-\left((T^{(n)})^2+2\Delta^{(n)}\right)K+T^{(n)}\Delta^{(n)} K^{3/2}}
\\
\end{eqnarray}
\end{widetext}
where $T^{(n)}$ and $\Delta^{(n)}$ are the trace and the determinant of the matrix $\boldsymbol{\bar{\mathcal{J}}}^{(n)}$\@.

The expansion of these expressions for large $K$ gives
\begin{eqnarray}\label{CorfullK}
\nonumber
\resizebox{\linewidth}{!}{$
C^{(n)}_{EE}=-\frac{A_E}{N}-\frac{1}{N\sqrt{K}}\frac{A_I(\bar{\mathcal{J}}_{EI}^{(n)})^2+A_E(\Delta^{(n)}+(\bar{\mathcal{J}}_{II}^{(n)})^2)}{T^{(n)}\Delta^{(n)}}+\mathcal{O}(\frac{1}{NK})
$}\\ \nonumber
\resizebox{\linewidth}{!}{$
C^{(n)}_{EI}=\frac{1}{N\sqrt{K}}\frac{A_E\bar{\mathcal{J}}^{(n)}_{II}\bar{\mathcal{J}}^{(n)}_{IE}+A_I\bar{\mathcal{J}}^{(n)}_{EE}\bar{\mathcal{J}}^{(n)}_{EI}}{T^{(n)}\Delta^{(n)}}+\mathcal{O}(\frac{1}{NK})
$}\\ \nonumber
\resizebox{\linewidth}{!}{$
C^{(n)}_{II}=-\frac{A_I}{N}-\frac{1}{N\sqrt{K}}\frac{A_E(\bar{\mathcal{J}}_{IE}^{(n)})^2+A_I(\Delta^{(n)}+(\bar{\mathcal{J}}_{EE}^{(n)})^2)}{T^{(n)}\Delta^{(n)}}+\mathcal{O}(\frac{1}{NK})
$}\\
\end{eqnarray}
Thus in general when $K$ is large $C^{(n)}_{EE}$ and $C^{(n)}_{II}$ are very small, namely $\mathcal{O}(1/N)$,  with negative prefactors which do not depend on $K$\@. As for $C^{(n)}_{EI}$, it is smaller than $C^{(n)}_{EE}$ and $C^{(n)}_{II}$ by a factor $1/\sqrt{K}$\@. 

It should be noted that to derive Eq.~\eqref{CorfullK} we assumed that  $T^{(n)}\neq0$ and $\Delta^{(n)}\neq0$. Equation \eqref{Corfull} indicates, however, that when $T^{(n)}=0$ and $\Delta^{(n)}\neq0$, or $T^{(n)}\neq0$ 	and  $\Delta^{(n)}=0$,  $C^{(n)}_{EE}$, $C^{(n)}_{II}$ and $C^{(n)}_{EI}$ are $\mathcal{O}(1/N)$\@.

The situation is different if $T^{(n)}=0$ {\it and} $\Delta^{(n)}=0$. Equation \eqref{Corfull} shows that in this case it is possible to get correlations which are $\mathcal{O}(K/N)$\@.

In the rest of this section we consider in detail two-population networks in which for the probabilities of connection only the first two Fourier modes are non-zero
\begin{equation}\label{cosineprof}
P^{\alpha\beta}_{ij}=\frac{K}{N}\left[1+2f^{(1)}_{\alpha\beta}\cos\left(\theta_{i}^\alpha-\theta_{j}^\beta\right)\right], 
\end{equation}
with $\alpha=E,I, \beta=E,I$.

\subsection{$T^{(1)} \neq 0$ and $\Delta^{(1)} \neq 0$} \label{SS2popGeneral}

For $f^{(1)}_{\alpha\beta}$ such that  $T^{(1)} \neq 0$ and $\Delta^{(1)} \neq 0$  we have in the large $N, K$ limit 
\begin{eqnarray}\label{CorInAll}
C_{EE}(\Delta)&=&-\frac{A_E}{N} (1+2\cos{\Delta})+\mathcal{O}(\frac{1}{{N}\sqrt{K}})\\ \nonumber
C_{EI}(\Delta)&=& \frac{1}{N\sqrt{K}}(\bar{C}^{(0)}_{EI}+2\bar{C}^{(1)}_{EI}\cos(\Delta))+\mathcal{O}(\frac{1}{NK}) \\ \nonumber
C_{II}(\Delta)&=&-\frac{A_I}{N}(1+2\cos{\Delta})+\mathcal{O}(\frac{1}{N\sqrt{K}}) \\ \nonumber
\end{eqnarray}
where $\bar{C}^{(n)}_{EI} \equiv N\sqrt{K} C^{(n)}_{EI}$ are $\mathcal{O}(1)$. Thus, in that limit, the spatial average and modulation of the correlations within the E and I populations do not depend on $K$, at the leading order. Moreover, $C_{EE}(\Delta)$ and $C_{II}(\Delta)$ depend on the synaptic strengths only because the autocorrelations $A_E$ and $A_I$ depend on these parameters.  

\begin{figure}[htp]
\begin{center}
\includegraphics[clip,width=0.5\textwidth]{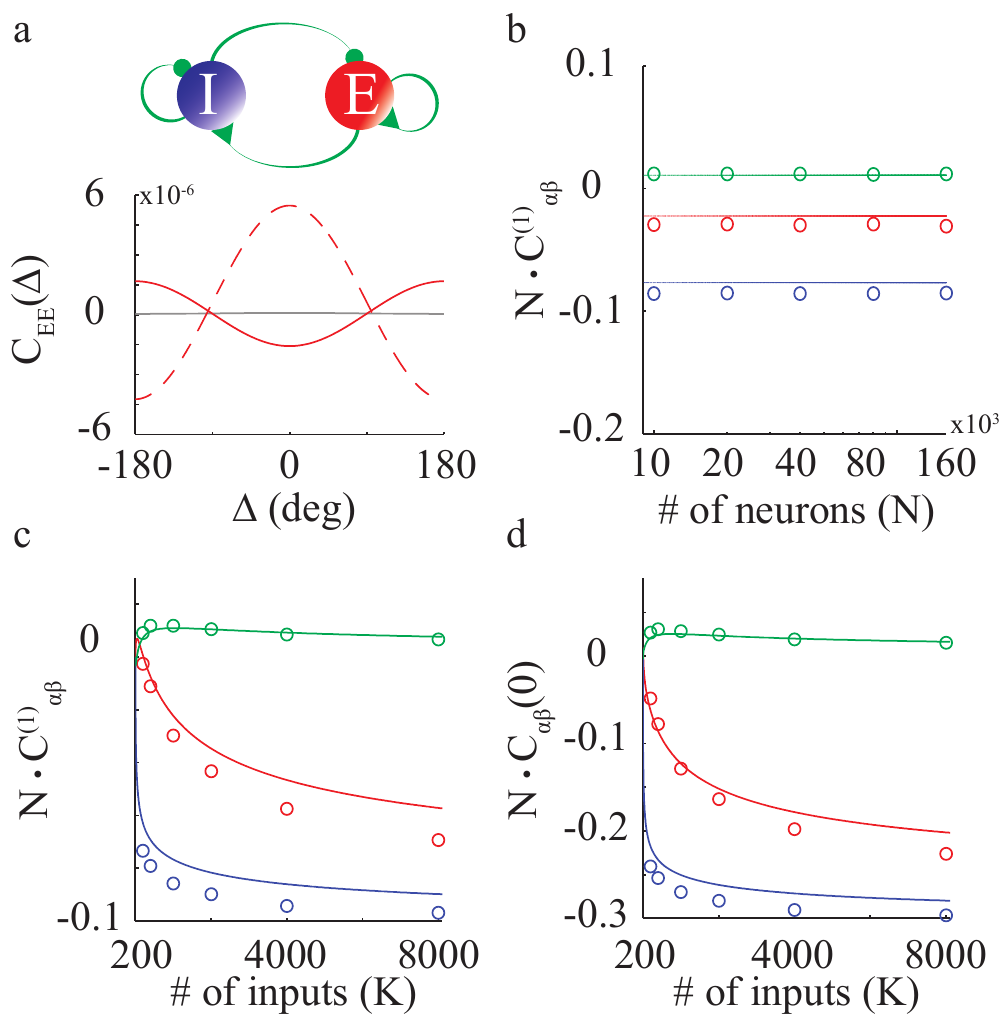}
\end{center}
\caption{{\bf Correlations in two-population E-I networks.} Connection probabilities are as in Eq.~\eqref{cosineprof}. Panels a, c, d: $N=40000$. {\bf a.} Top panel: The network architecture. All connections are spatially modulated ($f^{(1)}_{\alpha\beta}\neq 0$). Bottom panel: Simulation results for $C_{EE}(\Delta)$ with $K=2000$. Red: $f^{(1)}_{EE}=f^{(1)}_{EI}=f^{(1)}_{IE}=f^{(1)}_{II}=0.25$. For comparison the correlation is also plotted for $f^{(1)}_{EE}=f^{(1)}_{EI}=f^{(1)}_{IE}=f^{(1)}_{II}=0$ (Gray). For red-solid  and gray lines other parameters are  $J_{EE}=0.3, J_{IE}=3, J_{EI}=2.5, J_{II}=5; I_{E}=0.3, I_{I}=0.3, T_{E}=1, T_{I}=0.7$. With these parameters $m_{E}^{(0)}\simeq 0.12, m_{I}^{(0)}\simeq 0.13$, $g_E\simeq 0.22,g_I\simeq 0.1$ and $A_E\simeq0.1, A_I\simeq 0.1$. For red-dashed line other vparameters are: $J_{EE} = 1;  J_{IE} = 2;  J_{EI} =
1;  J_{II} = 1.5;  I_E = 0.2;  I_I = 0.08$.  {\bf b.} $N C_{\alpha\beta}^{(1)}$ vs. $N$ for $K=1000$. {\bf c.} $N C_{\alpha\beta}^{(1)}$ vs. $K$. {\bf d.} $N C_{\alpha\beta}(0)$ vs. $K$  ($N=40000$). In panels b, c, d:  Solid line: Analytics, Eq.~\eqref{Corfull}. Circles: Simulations. Red: $\alpha=\beta=E$. Blue: $\alpha=\beta=I$. Green: $\alpha=E, \beta=I$. Parameters are as for red-solid line in panel a.
}
\label{fig:BinaryAll}
\end{figure}

Figure~\ref{fig:BinaryAll} depicts simulation results for $N=40000$ and $K=2000$. Figure \ref{fig:BinaryAll}a plots $C_{EE}(\Delta)$ for $f^{(1)}_{\alpha\beta}=0.25$ ($\alpha, \beta \in {E,I}$) and two sets of values for the interaction strengths (solid and dashed lines). For comparison we also plots the results of a simulation when the connectivity is unstructured  ($f^{(1)}_{\alpha\beta}=0; \alpha, \beta \in {E,I} $; gray line). In all these cases $C_{EE}(\Delta)$ is very small (note the scale on thr y-axis). When the connectivity is spatially modulated $C_{EE}(\Delta)$ varies with distance. However, the spatial averages are comparable in the three cases considered ($\frac{1}{2\pi}\int C_{EE}(\Delta)d\Delta\sim -0.2 \times 10^{-5}$). This is in agreement with Eq.~\eqref{CorInAll} since, to the leading order, the auto-correlations, $A_E$ and $A_I$, are not expected to depend on whether the connectivity is spatially modulated or not.

Note that according to  Eq.~\eqref{CorInAll}, $C_{EE}(\Delta)$ and $C_{II}(\Delta)$ are negative for close-by neurons ($\Delta$ small) and positive for neurons that are far apart. This is the case for the set of parameters corresponding to the solid line in Fig.~\ref{fig:BinaryAll}a (see also Fig.~\ref{fig:BinaryAll}b-d). However, when finite $K$ corrections are not negligible, which happens when $|T^{(1)}\Delta^{(1)}|$ is sufficiently small, short range correlations can be positive and longer range correlations negative (Eq.~\eqref{CorInAll}). This is the case in Fig.~\ref{fig:BinaryAll}a, dashed line.

Figure~\ref{fig:BinaryAll}b compares simulations (circles) and analytical results (solid lines) 
for the dependence on $N$ of $N C_{\alpha \beta}^{(1)}$ (parameters as in panel a, red solid line). It shows that the spatial modulation of the correlations in the simulation is close to the large $K$ analytical results. Figure~\ref{fig:BinaryAll}c-d depict the dependence of $N C_{\alpha \beta}$ on $K$. In the whole range of $K$ considered here simulations and analytical results are close. For  $|C_{\alpha \alpha}|$ the nearby correlations and modulation amplitudes increase with $K$ and are much larger than that of $|C_{EI}|$.

\subsection{$T^{(1)}=\Delta^{(1)}=0$}\label{SsFF}

\begin{figure}[htp]
\begin{center}
\includegraphics[clip,width=0.5\textwidth]{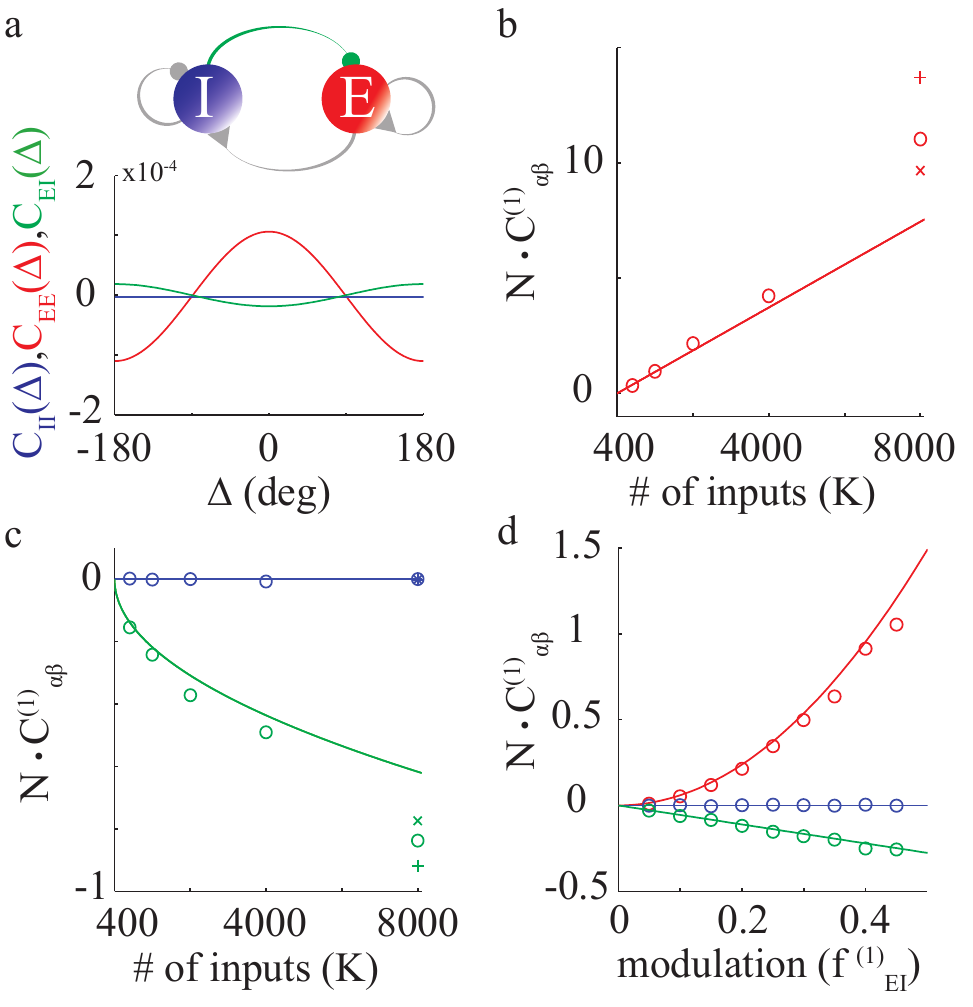}
\end{center}
\caption{{\bf Correlations in a two population network with an explicit feedfworward structure.}  Connection probabilities as in Eq.~\eqref{cosineprof}. Parameters:   $f^{(1)}_{EE}=f^{(1)}_{IE}=f^{(1)}_{II}=0$. Panels a, b, c:  $f^{(1)}_{EI}=0.25$. Interaction strengths as in Fig.~\ref{fig:BinaryAll}. {\bf a.} Top panel: The network architecture. Green: the connections which are spatially modulated ($f^{(1)}_{EI}\neq 0$). Connections in gray are unstructured ($f^{(1)}_{\alpha\beta}= 0$). Bottom panel: Simulation results for $C_{\alpha \beta}(\Delta)$;  $N=40000$, $K=2000$ .{\bf b, c.} $N C_{\alpha\beta}^{(1)}$ vs. $K$. Solid lines: Analytical results, Eq.~\eqref{CorInE}. Simulation results are plotted for $N=20000$ (plus), $N=40000$ (circles) and $N=80000$ (crosses). {\bf b.} $N C_{EE}^{(1)}$. {\bf c.} Blue; $N C_{II}^{(1)}$. Green:  $N C_{EI}^{(1)}$. {\bf d.} $N C_{\alpha\beta}^{(1)}$ vs. $f^{(1)}_{EI}$. Solid lines: Analytical results, Eq.~\eqref{CorInE}; $N=40000$, $K=400$. 
}
\label{fig:BinaryEI}
\end{figure}

We now consider a network in which $f^{(1)}_{EI}\neq 0$ and $f^{(1)}_{EE}=f^{(1)}_{II}=f^{(1)}_{IE}=0$. The spatially modulated component of the interaction has therefore an explicit feedforward structure (Fig.~\ref{fig:BinaryEI}a, top panel) and  $T^{(1)}=\Delta^{(1)}=0$.

Solving Eq.~\eqref{Corfull} shows that the correlations are on average $\mathcal{O}(1/N)$ and that their modulations are
\begin{eqnarray}\label{CorInE}
C_{EE}^{(1)}&=&\frac{K}{N}\frac{A_I}{2} |\bar{\mathcal{J}}_{EI}^{(1)}|^2\\ \nonumber
C_{EI}^{(1)}&=&-\frac{\sqrt{K}}{N}\frac{A_I}{2} |\bar{\mathcal{J}}_{EI}^{(1)}|\\ \nonumber
C_{II}^{(1)}&=& 0\\ \nonumber
\end{eqnarray}
As a result, correlations in the E population are spatially modulated and $\mathcal{O}(K/N)$.  They are positive for short range and negative for long range. This is in contrast to the correlations in the inhibitory population which are not spatially modulated and $\mathcal{O}(1/N)$ while the $EI$ correlations are spatially modulated and $\mathcal{O}(\sqrt{K}/N)$. 

Figure \ref{fig:BinaryEI} compares analytical results with simulations. Panel a plots simulation results for $C_{\alpha \beta}(\Delta)$ for $f_{EI}^{(1)}=0.25$. Other parameters are as in Fig.~\ref{fig:BinaryAll}a (grey solid line). Thus, the locally averaged firing rates are the same as in simulations in the latter figure. Correlations in the inhibitory neurons (blue) are extremely weak and are not spatially modulated. In contrast, the correlations of excitatory pairs (red) are larger by two orders of magnitude compared to those in Fig.~\ref{fig:BinaryAll}. Correlations between E and I neurons (green) are weaker than those between E neurons. They are  negative for nearby neurons while for nearby excitatory pairs they are positive. All these features are in agreement with our analytical results, 

The spatial modulation of the EE correlations increases with $K$ (Fig.~\ref{fig:BinaryEI}b, circles). There is quantitative agreement between simulations and theory (Eq.~\eqref{CorInE})  up to $K\approx4000$ for $N=40000$. In this range, $C_{EE}^{(1)}$ in the simulations varies linearly with $K$. For larger values of $K$, $C_{EE}^{(1)}$ is larger than predicted by Eq.~\eqref{CorInE}. This is because this equation was derived by linearizing the dynamics, which is only valid  when correlations are not too large. In fact, simulation results for fixed $K$ deviate less from Eq.~\eqref{CorInE} when $N$ is increased (Fig.\ref{fig:BinaryEI}b-c). When $K$ increases, $C_{EI}^{(1)}$, becomes more negative (Fig.~\ref{fig:BinaryEI}c; green). Here too, for $N=40000$ simulations agree well with Eq.~\eqref{CorInE} up to $K \approx 4000$ and deviations are smaller when $N$ is larger. The spatial modulation of the II correlations in the simulations are extremely small (Fig.~\ref{fig:BinaryEI}c; blue) as the theory predicts. 

Finally, according to Eq.~\eqref{CorInE}, the spatial modulation, $C_{EE}^{(1)}$, increases quadratically with $f^{(1)}_{EI}$ whereas  $C_{EI}^{(1)}$ varies linearly with this parameter. Our simulations are in very good  agreement with these analytical results (Fig.~\ref{fig:BinaryEI}d).

\subsection{$\Delta^{(1)}=0$, $T^{(1)}\neq 0$}

\begin{figure}[htp]
\begin{center}
\includegraphics[clip,width=0.5\textwidth]{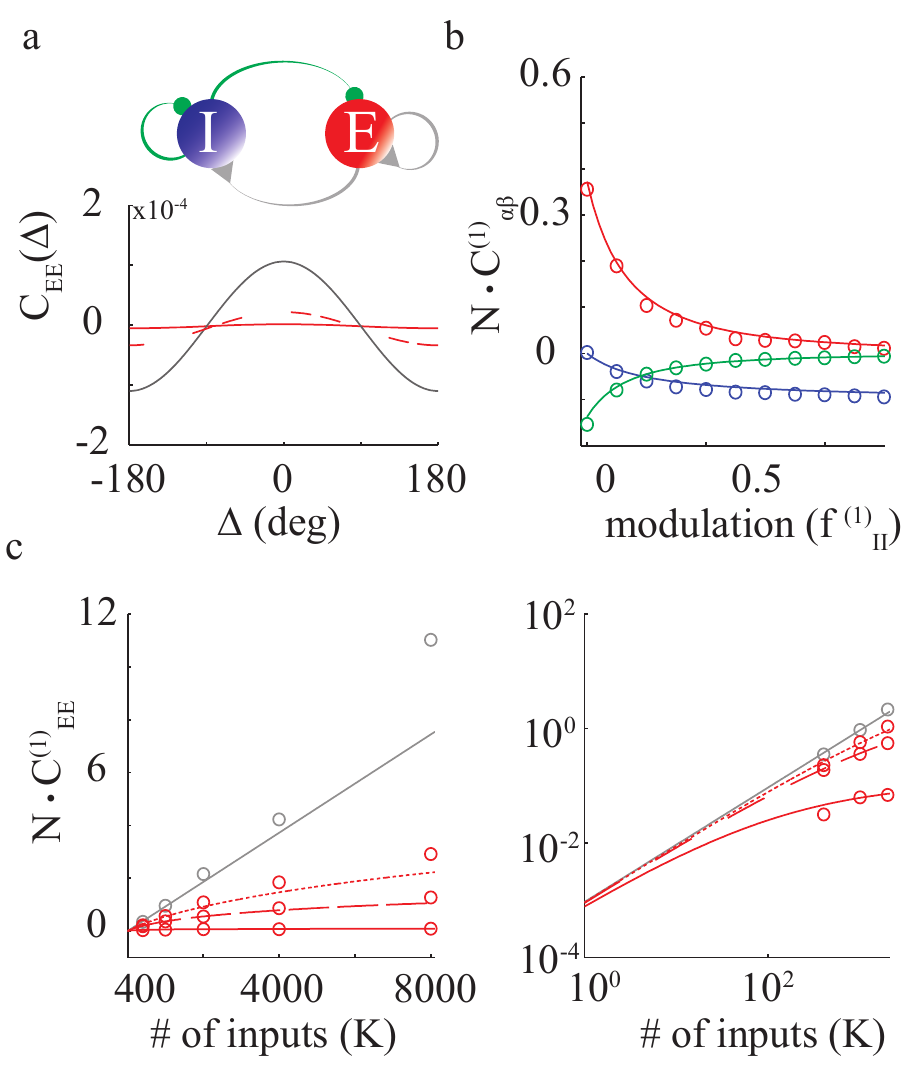}
\end{center}
\caption{
{\bf Spatial modulation in the II interactions suppresses the correlations in the E populations.} Same network as in Fig.~\ref{fig:BinaryEI} but  $f^{(1)}_{II}=0.25$. $N=40000$. {\bf a.} Top panel: The network architecture. Interactions plotted in green are spatially modulated. Bottom panel: Simulation results for $C_{EE}(\Delta)$. $N=40000$, $K=2000$. Gray line: $f^{(1)}_{II}=0$; Dashed- red: $f^{(1)}_{II}=0.05$; Solid-red : $f^{(1)}_{II}=0.25$. {\bf b.} $N C_{\alpha\beta}^{(1)}$ vs. $f^{(1)}_{II}$. Circles: Simulations. Solid lines: Solution of Eq.~\eqref{CorEI-II}. $N=40000$, $K=400$. Colors are as in Fig.~\ref{fig:BinaryEI}. {\bf c.} Left panel: $N C_{EE}^{(1)}$ vs. $K$. Solid lines: Eq.~\eqref{CorEI-II}. Circles: Simulations.  Gray line: $f^{(1)}_{II}=0$; Dotted red: $f^{(1)}_{II}=0.025$; Dashed red: $f^{(1)}_{II}=0.05$; Solid red : $f^{(1)}_{II}=0.25$. Right panel: Same as left panel but in a log-log scale.  }
\label{fig:BinaryEI-II} 
\end{figure}

The network investigated in \ref{SsFF} has an explicit feedforward structure. Adding any spatial modulation to the II connectivity ($f^{(1)}_{II} \neq 0$, top panel of Fig.~\ref{fig:BinaryEI-II}a) destroys this structure and now $T^{(1)}\neq 0$ (while $\Delta^{(1)}=0$). Solving  Eq.~\eqref{CorrEqMat} for that case, one finds:
\begin{eqnarray}\label{CorEI-II}
C_{EE}^{(1)}&=&\frac{K}{N}A_I\frac{ |\bar{\mathcal{J}}_{EI}^{(1)}|^2}{(2+\sqrt{K}|\bar{\mathcal{J}}_{II}^{(1)}|)(1+\sqrt{K}|\bar{\mathcal{J}}_{II}^{(1)}|)}\\ \nonumber
C_{EI}^{(1)}&=&-\frac{K}{N}A_I\frac{|\bar{\mathcal{J}}_{EI}^{(1)}|}{\sqrt{K}(2+\sqrt{K}|\mathcal{J}_{II}^{(1)}|)(1+\sqrt{K}|\mathcal{J}_{II}^{(1)})|}\\ \nonumber
C_{II}^{(1)}&=&-\frac{K}{N}A_I\frac{ |\bar{\mathcal{J}}_{II}^{(1)}|}{\sqrt{K}(1+\sqrt{K}|\bar{\mathcal{J}}_{II}^{(1)}|)}\\ \nonumber
\end{eqnarray}
In the large $N, K$ limit, $C_{EE}(\Delta)$ and $C_{II}(\Delta)$ are both $\mathcal{O}(1/N)$ and do not depend on $K$, whereas $C_{EI}(\Delta)$ is $\mathcal{O}(1/(N\sqrt{K}))$. Therefore the addition of a spatial modulation in the II interactions suppresses the correlations that inhibitory projections induce in the E population.

To understand further the origin of this, let us consider the quenched average correlations of the inputs, $Q_{\alpha \beta}(\Delta)$ (Eq.~\eqref{CorInputsDef}). Using Eq.~\eqref{CorInputs}, one finds
\begin{eqnarray*}
Q^{(1)}_{EE}&=&K(\bar{\mathcal{J}}^{(1)}_{EI})^2( C^{(1)}_{II}+\frac{A_{I}}{N})\\
Q^{(1)}_{EI}&=&K \bar{\mathcal{J}}^{(1)}_{EI} \bar{\mathcal{J}}^{(1)}_{II}(C^{(1)}_{II}+\frac{A_{I}}{N})\\
Q^{(1)}_{II}&=&K(\bar{\mathcal{J}}^{(1)}_{II})^2( C^{(1)}_{II}+\frac{A_{I}}{N})\\
\end{eqnarray*}
When $\bar{\mathcal{J}}^{(1)}_{II}= 0$,  $C^{(1)}_{II}=0$, therefore $Q^{(1)}_{EE} = \frac{K}{N}(\bar{\mathcal{J}}^{(1)}_{EI})^2A_{I}$. On the other hand, when $\bar{\mathcal{J}}^{(1)}_{II}\neq 0$,  $C^{(1)}_{II}$ is given by Eq.\eqref{CorEI-II}. This yields
\begin{equation}\label{inputE}
Q^{(1)}_{EE}=\frac{A_{I}}{N}\frac{K (\bar{\mathcal{J}}^{(1)}_{EI})^2}{1+\sqrt{K}|\bar{\mathcal{J}}^{(1)}_{II}|}
\end{equation}\@
Thus, when II interactions are spatially modulated, a cancellation between terms which are $\mathcal{O}(K/N)$, reduces $Q^{(1)}_{EE}$ by a factor of $\mathcal{O}(1/\sqrt{K})$. As a result, $C^{(1)}_{EE}$ is much smaller than when II interactions are not modulated. A similar argument explains the suppression in $C^{(1)}_{EI}$.

Fig.~\ref{fig:BinaryEI-II}a depicts simulation results for $f_{EI}^{(1)}=0.25$ and three values of $f_{II}^{(1)}$. It demonstrates the suppression of correlations in the excitatory population when II interactions are also modulated. The dependence of this effect on $f_{II}^{(1)}$ is depicted in Fig.~\ref{fig:BinaryEI-II}b. Increasing $f^{(1)}_{II}$, decreases the modulation of all the correlations in very good agreement with the analytical results (compare circles and solid lines).

According to Eq.~\eqref{CorEI-II}, the correlation in the E population always increases linearly with $K$, for small $K$\@. The cross-over between this regime, where $C_{EE}(\Delta)$ is $\mathcal{O}(K/N)$, and the large $K$ regime, where $C_{EE}(\Delta)$ is $\mathcal{O}(1/N)$,  occurs for $ \sqrt{K} \simeq 1/\bar{\mathcal{J}}_{II}^{(1)}$\@. Figure \ref{fig:BinaryEI-II}c depicts this crossover in numerical simulations. Thus, although in the large $N, K$ limit a transition from $\mathcal{O}(K/N)$ to $\mathcal{O}(1/N)$ correlations occurs as soon as $f^{(1)}_{II}\neq 0$, this qualitative difference is significant only when $K$ is sufficiently large. In other words, for moderately large number of inputs per neuron, correlations can exhibit a close to linear increase even if the structure of spatial modulation of the interaction matrix is not completely feedforward.

\section{Scaling correlation theorems}\label{Stheorem}

In the previous section we studied networks with two neuronal populations. In this case, it is straightforward to analytically derive explicit expressions for the correlations. These expressions are simple enough to fully classify how the network structure affects the scaling (with $K$ and $N$) of the correlations. For networks consisting of more than two populations, analytical expressions for the correlations can in principle be derived. However, dealing with these expressions becomes rapidly impractical as the number of populations increases. In the following we adopt an alternative approach. We prove two general theorems, which, given the network architecture, allow us to determine how correlations vary with $K$, without computing 
these correlations explicitly.

To prove these theorems, we rewrite Eq.~\eqref{CorrEqMat} in the Jordan basis of $\boldsymbol{\bar{\mathcal{J}}}^{(n)}$. We write
\begin{eqnarray}\label{TransJordan}
\boldsymbol{\bar{\mathcal{J}}}^{(n)}&=&\boldsymbol{U}^{(n)}\boldsymbol{\mathcal{J}}^{(n)}_{\jor}[\boldsymbol{U}^{(n)}]^{-1}\\ \nonumber
[\boldsymbol{\bar{\mathcal{J}}}^{(n)}]^{\dagger}&=&\boldsymbol{V}^{(n)}[\boldsymbol{\mathcal{J}}_{\jor}^{(n)}]^{*}[\boldsymbol{V}^{(n)}]^{-1}
\end{eqnarray}

Here $x^{*}$ denotes the complex conjugate of $x$, $\boldsymbol{U}^{(n)}$ ($\boldsymbol{V}^{(n)}$) are matrices whose rows are the generalized eigenvectors of $\boldsymbol{\bar{\mathcal{J}}}^{(n)}$, and $\boldsymbol{\mathcal{J}}_{\jor}^{(n)}$ is the Jordan normal form of $\boldsymbol{\bar{{\mathcal{J}}}}^{(n)}$. This implies that $[\boldsymbol{\mathcal{J}}_\jor^{(n)}]^{*}$ is the Jordan form of $[\boldsymbol{\bar{\mathcal{J}}}^{(n)}]^{\dagger}$ is the 
$[{\boldsymbol{\bar{\mathcal{J}}}}^{(n)}]^{\dagger}$. 
We can write
\[
[\boldsymbol{\mathcal{J}}^{(n)}_{\jor}]_{\mu\nu}=\lambda^{(n)}_{\mu}\delta_{\mu\nu}+\epsilon^{(n)}_{\mu}\delta_{\mu,\nu-1}
\]
with $\mu, \nu \in {1,...,D}$ and $\epsilon^{(n)}_{\mu}=1$ within a Jordan block and zero otherwise. 

Equation \eqref{CorrEqMat} then yields
\begin{widetext}
 \begin{eqnarray}\label{JordanForm}
 \left[2-\sqrt{K}(\lambda^{(n)}_{\mu}+[\lambda_{\nu}^{(n)}]^{*})\right]\hat{C}^{(n)}_{\mu\nu}&=&\sqrt{K}(\epsilon_{\mu}^{(n)}\hat{C}^{(n)}_{\mu+1,\nu}+\epsilon^{(n)}_{\nu-1}\hat{C}^{(n)}_{\mu,\nu-1} )\\\nonumber
 &+&\frac{\sqrt{K}}{N}(\epsilon_{\mu}^{(n)}\hat{A}^{(n)}_{\mu+1,\nu}+\epsilon^{(n)}_{\nu-1}\hat{A}^{(n)}_{\mu,\nu-1} )+\frac{\sqrt{K}}{N}(\lambda_{\mu}^{(n)}+[\lambda_{\nu}^{(n)}]^{*})\hat{A}^{(n)}_{\mu\nu}\\ \nonumber
\end{eqnarray}
\end{widetext}
where we have defined
\begin{eqnarray}\label{Trans}
\boldsymbol{\hat{C}}^{(n)}&=&[\boldsymbol{U}^{(n)}]^{-1}\boldsymbol{C}^{(n)}\boldsymbol{V}^{(n)}\\ \nonumber
\boldsymbol{\hat{A}}^{(n)}&=&[\boldsymbol{U}^{(n)}]^{-1}\boldsymbol{A}^{(n)}\boldsymbol{V}^{(n)}
\end{eqnarray}
Note that while the matrix $\boldsymbol{C}^{(n)}$ is symmetric and the matrix $\boldsymbol{A}^{(n)}$ is diagonal, this is not in general the case for $\boldsymbol{\hat{C}}^{(n)}$ and $\boldsymbol{\hat{A}}^{(n)}$.

Let us assume that the network is in a stable balanced state in which the matrix $\boldsymbol{\hat{A}}^{(n)}$ has no zero elements.  
In Appendix \ref{ApProof} we prove

\medskip

\noindent {\it Correlation Theorem 1:}  The $n^{th}$ Fourier mode of the correlation matrix scales as $\mathcal{O}(1/N)$ if and only $\boldsymbol{\mathcal{J}}_{\jor}^{(n)}$ does not have a Jordan block whose real part is a shift matrix (A shift matrix, $\boldsymbol{S}$, of dimension $P$ is a $P \times P$ matrix of the form $S_{\mu\nu}=\delta_{\mu+1,\nu}$).

\medskip

\noindent {\it Correlation Theorem 2:} If $\boldsymbol{\bar{\mathcal{J}}}^{(n)}$ has at least one Jordan block whose real part is a shift matrix, the $n^{th}$ Fourier mode of the correlation matrix is $\mathcal{O}(K^{P(n)-1}/N)$, where $P(n)$ is the dimension of the largest block in $\boldsymbol{\mathcal{J}}^{(n)}_{\jor}$, whose real part is a shift matrix.

\medskip

\noindent{\it Corollary:} The $n^{th}$ Fourier mode of the correlation matrix is $\mathcal{O}(K^{D-1}/N)$ if and only if $\boldsymbol{\mathcal{J}}^{(n)}$ is nilpotent of degree $D$. 

\medskip
In Appendix  \ref{ApProof} we also show how to extend these resukts to case where $\boldsymbol{\hat{A}}^{(n)}$ has zero elements.

The matrix $\boldsymbol{U}^{(n)}$ can be viewed as a transformation of the original network of $D$ populations into a network of $D$ effective populations. The condition that $\boldsymbol{\bar{\mathcal{J}}}^{(n)}$ has a Jordan block which is a shift matrix of size $P$, can be interpreted as the existence of $P$ effective populations whose effective interaction matrix is feedforward in its $n^{th}$ Fourier mode. In other words, the original network exhibits a  hidden feedforward structure, which is embedded in the $n^{th}$ Fourier mode of its connectivity. Theorems 1 and 2 therefore implies that only when such a structure exists, the $n^{th}$ mode of the correlation matrix increases with $K$. To know which elements in this  matrix increase with $K$ one has to compute the matrices $\boldsymbol{U}^{(n)}$ and $\boldsymbol{V}^{(n)}$ (see Eq.~\eqref{Trans}).

\section{Applications of the correlation theorems}\label{AppliTh}

In this section we consider networks comprising $D$ populations with connection probabilities
\begin{equation}\label{fab2}
P_{ij}^{\alpha\beta}=\frac{K}{N}\left(1+2f_{\alpha \beta}^{(1)}\cos(\theta_i^{\alpha}-\theta_j^{\beta})\right)
\end{equation}
with $\alpha=1,...,D$ and  $\beta=1,...,D$.

\subsection{Two population networks}

For a network of two populations the Jordan form of the matrix $\boldsymbol{\bar{\mathcal{J}}}^{(1)}$ has the form:
$\boldsymbol{\mathcal{J}}^{(1)}_\jor= \left( \begin{array}{cc}
÷
\lambda_1 & 0 \\
0& \lambda_2\\
\end{array} \right)$
if $\boldsymbol{\bar{\mathcal{J}}}^{(1)}$ is diagonalizable. Otherwise, it has the form:
$\boldsymbol{\mathcal{J}}^{(1)}_\jor= \left( \begin{array}{cc}
\lambda & 1 \\
0& \lambda\\
\end{array} \right)$
with $\lambda$ real. Theorem 1 and 2 imply that only in the second case with $\lambda=0$, some of the correlations $C_{\alpha \beta}(\Delta)$ are $\mathcal{O}(K/N)$. Otherwise, all correlations are $\mathcal{O}(1/N)$. It is equivalent to say that some correlations are $\mathcal{O}(K/N)$ if and only if $\boldsymbol{\bar{\mathcal{J}}}^{(1)} \neq \boldsymbol{0}$ and  $T^{(1)}=\Delta^{(1)}=0$. We therefore recover the result we derived in Section \ref{S2networks} without explicit computation of the correlations. 

As noted above, the correlation theorems do not tell us which of the elements of the correlation matrix are $\mathcal{O}(K/N)$ when $\boldsymbol{\mathcal{J}}^{(1)}_\jor= \left( \begin{array}{cc}
0 & 1 \\
0&  0\\
\end{array} \right)$. 
However, since a $2 \times 2$ matrix has such a Jordan form if and only if it is nilpotent ($\left(\boldsymbol{\bar{\mathcal{J}}}^{(1)}\right)^2=0$) we have to consider two types of connectivity:

Type 1: The network has an explicit feedforward structure, {\it i.e.}, the interaction matrix is either 
$\boldsymbol{\bar{\mathcal{J}}}^{(1)}=\left(\begin{array}{cc}
0 & \bar{\mathcal{J}}_{EI}^{(1)} \\
0 & 0 \\
\end{array} \right)$
\medskip
or
$\boldsymbol{\bar{\mathcal{J}}}^{(1)}=\left( \begin{array}{cc}
0 & 0 \\
\bar{\mathcal{J}}_{IE}^{(1)} & 0 \\
\end{array} \right)$.
\medskip

In the former case, the matrix $\boldsymbol{U}^{(1)}$ is $\boldsymbol{U}^{(1)} =\left(\begin{array}{cc}
1 & 0 \\
0 & 1/\bar{\mathcal{J}}_{EI}^{(1)} \\
\end{array} \right)
$, whereas  $\boldsymbol{V}^{(1)} = \left(\begin{array}{cc}
0 & 1 \\
1/\bar{\mathcal{J}}_{EI}^{(1)} & 0 \\
\end{array} \right)$. Then Eq.~\eqref{Trans} gives (see Appendix \ref{ApProof}, Eq.~\eqref{degen})
\begin{equation}\label{eScalingChat}
\boldsymbol{\boldsymbol{\hat{C}}}^{(1)}=\left(\begin{array}{cc}
\mathcal{O}(\sqrt{K}/N)& \mathcal{O}(K/N) \\
0 &\mathcal{O}(\sqrt{K}/N) \\
\end{array} \right)
\end{equation}
Using Eq.~\eqref{Trans}, one finds that $C_{EE}^{(1)}=\mathcal{O}(K/N)$, $C_{II}^{(1)}=0$,  whereas $C_{EI}^{(1)}=\mathcal{O}(\sqrt{K}/N)$, in agreement with Eq.~\eqref{CorInE}.
\medskip
A similar calculation in the latter case (when the modulation is in the IE interactions) gives $C_{EE}^{(1)}=0$, $C_{II}^{(1)}=\mathcal{O}(K/N)$, and $C_{EI}^{(1)}$ is $\mathcal{O}(\sqrt{K}/N)$, all in agreement with Eq.~\eqref{Corfull}.

\medskip

Type 2: $\boldsymbol{\bar{\mathcal{J}}}^{(1)}= c\left( \begin{array}{cc}
1 & -a \\
1/a & -1 \\
\end{array} \right)$
\medskip
where $a,c>0$. 

In this case the network has no {\it explicit} feedforward structure  since all four interactions ($EE, EI, IE, II$) are spatially modulated. It has, however, a {\it hidden} feedforward structure, as revealed by the Jordan form of the interaction matrix. 

For this network, the transformation matrices are $\boldsymbol{U}^{(1)} =\left(\begin{array}{cc}
a & a/c \\
1 & 0 \\
\end{array} \right) $
and $\boldsymbol{V}^{(1)} = \left(\begin{array}{cc}
-1 & -1/c \\
a & 0 \\
\end{array} \right)$. Using Eq.~\eqref{Trans} and Eq.~\eqref{eScalingChat}, it is clear that the transformation from $\boldsymbol{\boldsymbol{\hat{C}}}^{(1)}$ to $\boldsymbol{\boldsymbol{C}}^{(1)}$ mixes elements which are $0$ and $\mathcal{O}(\sqrt{K}/N)$, with those which are $\mathcal{O}(K/N)$. Thus, while in $\boldsymbol{\boldsymbol{\hat{C}}}^{(1)}$ only the element $\hat{C}_{12}^{(1)}$ is $\mathcal{O}(K/N)$, all the elements of the correlation matrix $\boldsymbol{C}^{(1)}$ are $\mathcal{O}(K/N)$. This is also in line with Eq.~\eqref{Corfull}. 

\begin{figure}[htp]
\begin{center}
\includegraphics[clip,width=0.5\textwidth]{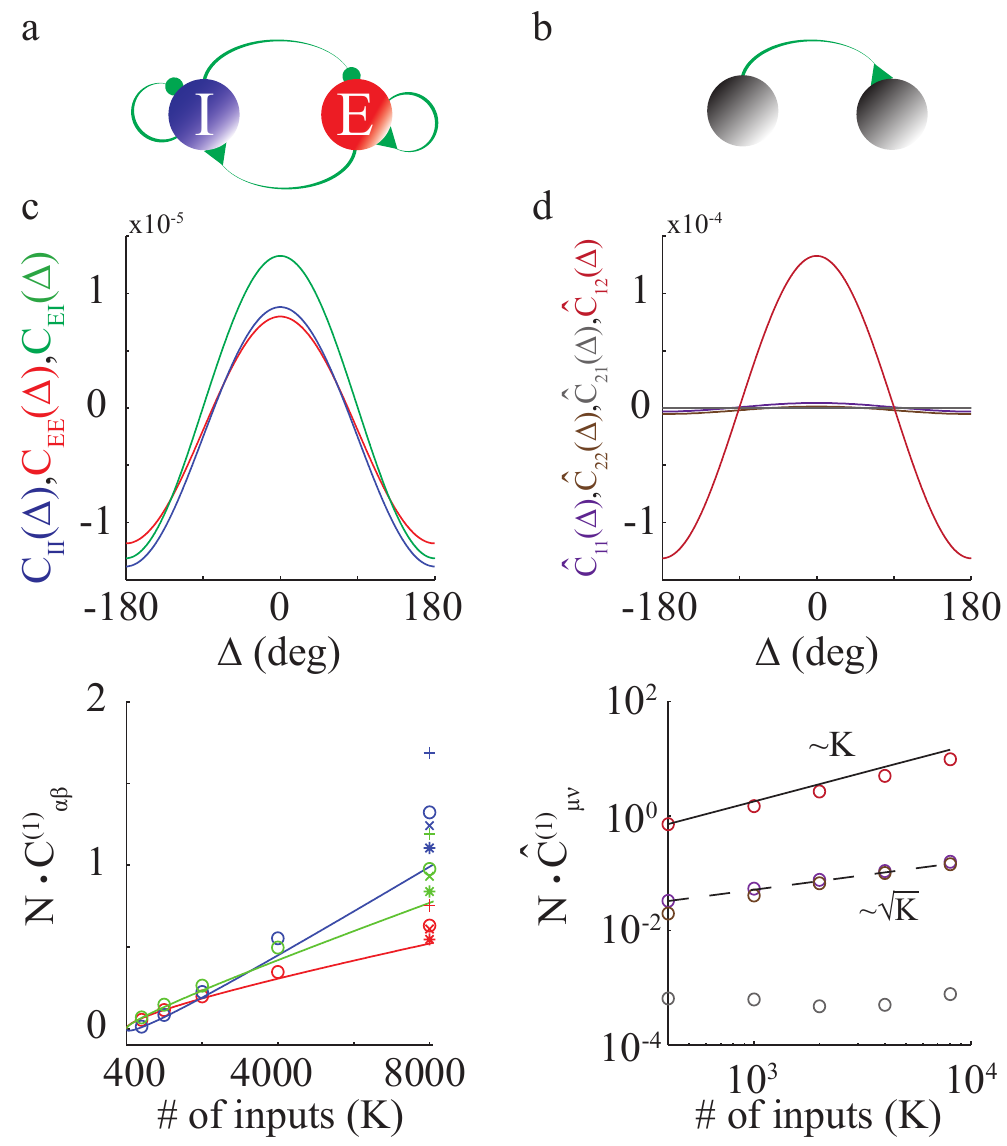}
\end{center}
\caption{{\bf Correlations in a two population network with a hidden feedforward structure.} {\bf a.} Network architecture. All connections are modulated. Connection probabilities are as in Eq.~\eqref{cosineprof} with $f^{(1)}_{EE}= 0.373, f^{(1)}_{EI}= 0.0224, f^{(1)}_{II}=0.0487, f^{(1)}_{IE}= 0.1623$. Connection strengths are as in Figs.~\ref{fig:BinaryAll}-\ref{fig:BinaryEI}. {\bf b.}  Hidden feedforward structure of the mode $n=1$ as revealed in the Jordan basis. {\bf c.} Top panel: Simulation results for $C_{\alpha \beta}(\Delta)$. $N=40000, K=2000$. Bottom panel: $N C_{\alpha\beta}^{(1)}$ vs. $K$. Solid lines: Analytical results (Eq.~\eqref{Corfull}). Simulations are plotted for $N=20000$ (plus), $N=40000$ (circles), $N=80000$ (cross) and $N=160000$ (asterisks). {\bf d.} Top panel: Correlation matrix in the Jordan basis of $\boldsymbol{\bar{\mathcal{J}}}^{(1)}$. Purple: $\hat{C}_{11}(\Delta)$; Brown: $\hat{C}_{22}(\Delta)$; Gray: $\hat{C}_{21}(\Delta)$; Red: $\hat{C}_{12}(\Delta)$. Bottom panel: $N \hat{C}_{\mu\nu}^{(1)}$ vs. $K$ in log-log scale. Solid line: Linear fit. Dashed line: Fit to a square root function.}
\label{fig:BinaryNilp}
\end{figure}

We consider an example of such a network in Fig.~\ref{fig:BinaryNilp}. The parameters $J_{\alpha \beta}$ and the external inputs are as in Fig.~\ref{fig:BinaryAll}-\ref{fig:BinaryEI-II}. Therefore, to leading order, the population averaged activities, $m_{\alpha}$, the autocorrelations, $A_{\alpha}$, and the population gains, $g_{\alpha}$ are the same as in Figs.~\ref{fig:BinaryAll}-\ref{fig:BinaryEI-II}. The modulation of the connection probability, $f^{(1)}_{EE},f^{(1)}_{IE}, f^{(1)}_{EI}, f^{(1)}_{II}$, are all non-zero (Fig.~\ref{fig:BinaryNilp}a) and tuned so that:
\[
\boldsymbol{\bar{\mathcal{J}}}^{(1)}=\frac{1}{20}\left( \begin{array}{cc}\label{hidden}
1 & -1/2 \\
2 & -1 \\
\end{array} \right)
\]
 
The Jordan form of this matrix is graphically represented in Fig.~\ref{fig:BinaryNilp}b. The top panel in Fig.~\ref{fig:BinaryNilp}c depicts simulation results for the correlations in  this network. They are all positive for close-by neurons and their spatial modulations increase approximately linearly with $K$ (Fig.~\ref{fig:BinaryNilp}c, bottom). The top panel of Fig.~\ref{fig:BinaryNilp}d shows that most of the power in these correlations results from the element $\hat{C}_{12}$ (red). The latter increases linearly with $K$ (Fig.~\ref{fig:BinaryNilp}d,  bottom), while  $\hat{C}_{11},\hat{C}_{22}= \mathcal{O}(\sqrt{K}/N)$ and $\hat{C}_{21}$ is two order of magnitude smaller and does not exhibit significant change with $K$ (note the log-log scale in this figure). These simulations are in quantitative agreement with Eq.~\eqref{Corfull} up to $K \approx 4000$ and the deviations for larger $K$ decrease with $N$ (as in Fig \ref{fig:BinaryEI-II}).

The interaction matrix, $\boldsymbol{\bar{\mathcal{J}}}^{(1)}$, depends on the matrix $\boldsymbol{{J}}^{(1)}$, and on the population gains,  $g_{E}$ and $g_{I}$. If the connectivity matrix has an explicit feedforward structure, the interaction matrix has also such a structure, independantly of the population gains. Thus, although changing the external inputs, $I_{E}, I_{I}$, modifies this gains, this does not destroy the feedforward structure and thus does not change the scaling  of the correlations with $K$ and $N$. In contrast, in networks with a hidden feedforward structure, this scaling is sensitive to perturbations in the external inputs since the hidden feedforward structure is destroyed unless the ratio $g_{E}/g_{I}$ remains constant.

\subsection{Examples with three populations or more}\label{SSeveralPop}

\begin{figure}[htp]
\begin{center}
\includegraphics[clip,width=0.5\textwidth]{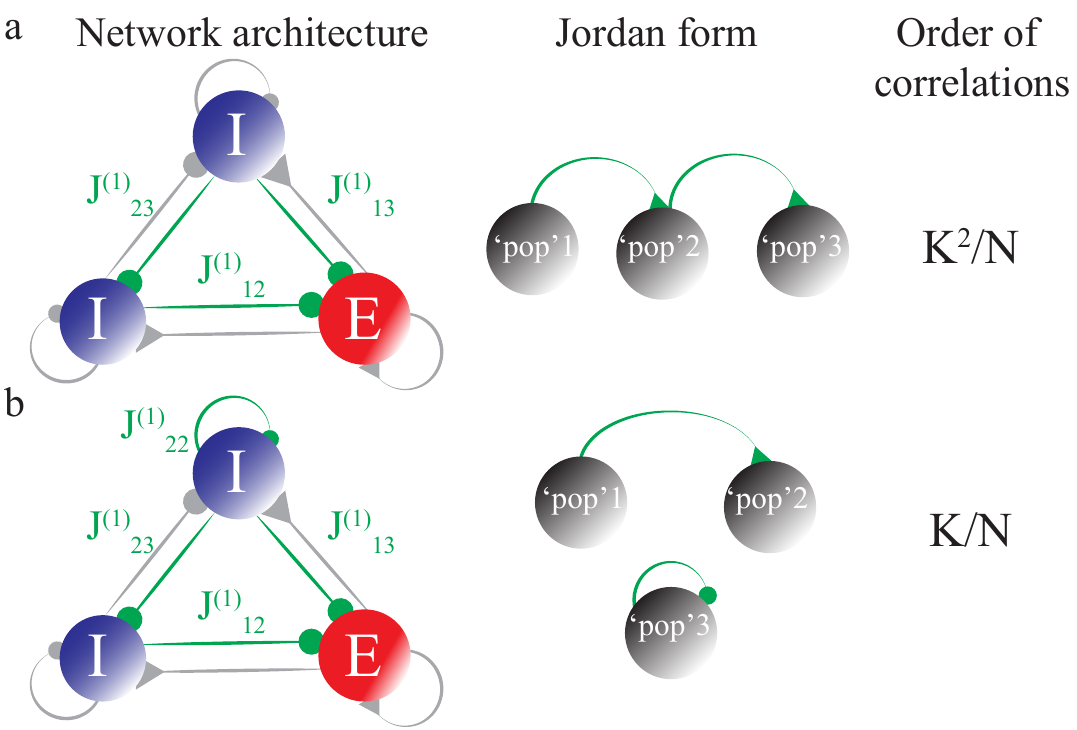}
\end{center}
\caption{{\bf Examples of networks with three populations and their Jordan representations.} Probability of connections are as in Eq.~\eqref{fab2}. {\bf a-b.} Left: A network of three populations, two inhibitory and one excitatory. Gray: Unstructured connections. Green: Spatially modulated connections. Middle: The Jordan representation of the $n=1$ Fourier mode of the network connectivity (left panel). Right: The scaling of the strongest correlations. {\bf b.} $J^{(1)}_{12},J^{(1)}_{13},J^{(1)}_{23}\neq0$. Other entries of the matrix $\boldsymbol{J}^{(1)}$ are zero (see main text). {\bf b.} Same as (a), but with $J^{(1)}_{22}\neq 0$.}
\label{fig:3Pop}
\end{figure}

The two networks depicted in Fig.~\ref{fig:3Pop} consist of one excitatory and two inhibitory populations. The positivity of the spatial averaged activity of these three populations, $m^{(0)}_{\alpha}$ ($\alpha=1,2,3$), constrains the parameters (see Section \ref{SSpatiotemporal}), $J_{\alpha\beta}$, through a set of inequalities. We leave the calculation of these conditions to the reader.

In both cases, the first Fourier mode of the population average connectivity matrix has the form
\begin{equation}\label{3by3}
\boldsymbol{{J}}^{(1)}=  \left( \begin{array}{ccc}
0 & J_{12}^{(1)} & J_{13}^{(1)} \\
0 & J_{22}^{(1)} & J_{23}^{(1)} \\
0 &  0 & 0  \\
\end{array} \right)
\end{equation}
with  $J^{(1)}_{12},J^{(1)}_{13},J^{(1)}_{23}< 0$ (left panels of Fig~\ref{fig:3Pop}a-c). 

In the network of Fig.~\ref{fig:3Pop}a, $J^{(1)}_{22}=0$. Therefore $\boldsymbol{{J}}^{(1)}$ and $\boldsymbol{\mathcal{J}}^{(1)}$ are nilpotent of degree $3$. According to the Corollary of Section \ref{Stheorem}, the first mode of the correlation matrix is $\mathcal{O}(K^2/N)$. 

The Jordan form of $\boldsymbol{\bar{\mathcal{J}}}^{(1)}$ is
\[
\boldsymbol{\mathcal{J}  }^{(1)}_{\jor} =  \left( \begin{array}{ccc}
0 & 1 & 0 \\
0 & 0 & 1 \\
0 &  0 & 0  \\
\end{array} \right).
\]
This is graphically represented in Fig.~\ref{fig:3Pop}a, middle panel. The matrix $\boldsymbol{\hat{C}}^{(1)}$ satisfies (see Appendix \ref{ApProof}, Eq.\eqref{degen} )
\[
\boldsymbol{\hat{C}}^{(1)}=  \left( \begin{array}{ccc}
\mathcal{O}(K/N) &  \mathcal{O}(K^{3/2}/N) & \mathcal{O}(K^{2}/N) \\
\mathcal{O}(K^{1/2}/N) & \mathcal{O}(K/N) & \mathcal{O}(K^{3/2}/N) \\
0 &  \mathcal{O}(K^{1/2}/N) & \mathcal{O}(K/N)  \\
\end{array} \right)
\]
Using the transformation matrices (Eq.\eqref{TransJordan}), one can show that correlations are $\mathcal{O}(K^2/N)$ only within the excitatory population. 

In the network in Fig.~\ref{fig:3Pop}b,  $J^{(1)}_{22}<0$. Therefore, the interaction matrix, $\boldsymbol{\bar{\mathcal{J}}}^{(1)}$, is not nilpotent. Its Jordan form is
\[
\boldsymbol{\mathcal{J}  }^{(1)}_{\jor} =  \left( \begin{array}{ccc}
0 &  1 & 0 \\
0 & 0 & 0 \\
0 &  0 & J_{22}^{(1)}  \\
\end{array} \right).
\]
The upper Jordan block is a shift matrix of degree 2. The corresponding feedforward structure is graphically represented in 
Fig.~\ref{fig:3Pop}b (middle panel). According to Theorem 2, $\boldsymbol{\hat{C}}^{(1)}$ is $\mathcal{O}(K/N)$.  It satisfies (see Appendix \ref{ApProof}, Eq.~\eqref{degen})
\[
\boldsymbol{\hat{C}}^{(1)}=  \left( \begin{array}{ccc}
\mathcal{O}(\sqrt{K}/N) &  \mathcal{O}(K/N) & \mathcal{O}(1/N) \\
0 & \mathcal{O}(\sqrt{K}/N) & \mathcal{O}(1/N) \\
\mathcal{O}(1/N) &  \mathcal{O}(1/N) & \mathcal{O}(1/N)  \\
\end{array} \right)
\]
and using the transformation matrices one can show that only $\boldsymbol{C}_{11}^{(1)}$ is $\mathcal{O}(K/N)$. Other correlations are either $\mathcal{O}(\sqrt{K}/N)$ or $\mathcal{O}(1/N)$.

\begin{figure*}[htp]
\begin{center}
\includegraphics[clip,width=\textwidth]{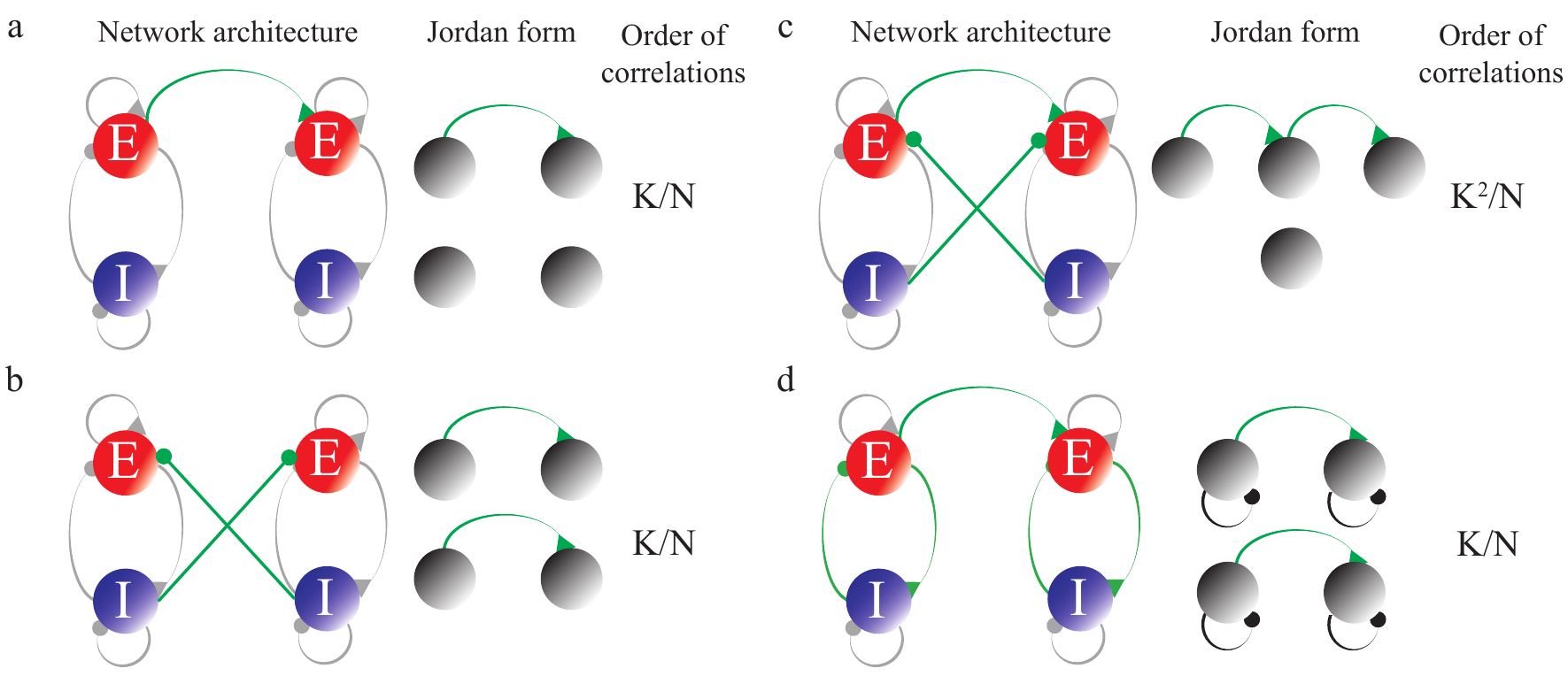}
\end{center}
\caption{{\bf Examples of networks with four populations and their Jordan representations.} Probability of connections are as in Eq.~\eqref{fab2}. {\bf a-c.} Left: The network consists of two coupled E-I networks. Gray: Unstructured connections. Green: Spatially modulated connections. Middle: The Jordan representation of the $n=1$ Fourier mode of the connectivity (network of the left panel. Right: The scaling of the strongest correlation. {\bf d.} Same as (a-c), with a population averaged connectivity matrix as in Eq.~\eqref{e4popExample}. Middle: The Jordan form is complex (Eq.~\eqref{eComplexJordan}). Black lines corresponds to complex eigenvalues.} 
\label{fig:4pop}
\end{figure*}

This approach can be generalized to arbitrary number of populations to classify the scaling of the correlation matrix for different architectures. Examples of networks with four populations are depicted in Fig~\ref{fig:4pop}, together with the graphic representations of their Jordan forms and the maximum order of the correlations.

\section{Constraint on scaling of the number of inputs with the network size}\label{SlargeK} 

In this section we assume that $K$ and $N$ scale together
\begin{equation}
K \propto N^{\gamma}
\end{equation}
with $0 <\gamma\leq 1$. 

Equation\eqref{CorrEqMat}, which determines the correlations of the neuronal activities, is obtained under the Ansatz that the correlations between the inputs $h_{i}^{\alpha}(t)$ and $h_{j}^{\beta}(t)$ are sufficiently small, namely $o(1)$ (see Appendix \ref{ApCorBinary}). This condition is more stringent than the balance correlation equation, Eq.~\eqref{BalanceCorr}. It can be written as

\begin{equation}\label{BalanceCorrStrong}
\boldsymbol{J}^{(n)}\boldsymbol{C}^{(n)}[\boldsymbol{J}^{(n)}]^{\dagger}=o(1/N^{\gamma})
\end{equation}
This condition constrains $\gamma$ as we now show. 

According to Theorem 1, if the Jordan form, $\mathcal{J}^{(n)}_{\jor}$, has no Jordan block whose real part is a shift matrix for any $n$, correlations will be $\mathcal{O}(1/N)$. This will also be the order of $\boldsymbol{J}^{(n)}\boldsymbol{C}^{(n)}[\boldsymbol{J}^{(n)}]^{\dagger}$. Therefore, Eq.~\eqref{BalanceCorrStrong} only requires $\gamma<\gamma_{\max}=1$.

If the Jordan form, $\boldsymbol{\mathcal{J}}^{(n)}_{\jor}$, contains a Jordan whose real part in the shift matrix, we have to apply Theorem 2. In this theorem, $P(n)$ is the dimension of the largest block in $\boldsymbol{\mathcal{J}}^{(n)}_{\jor}$, with a real part which is a shift matrix (see Section \ref{Stheorem}). Let us denote by $P_{\max}$ the largest $P(n)$ over all Fourier modes, i.e.,
\begin{equation}
P_{\max}=\max_n P(n)
\end{equation}

Equation \eqref{BalanceCorrStrong} implies
\[
\boldsymbol{\bar{\mathcal{J}}}^{(n)}\boldsymbol{C}^{(n)}[\boldsymbol{\bar{\mathcal{J}}}^{(n)}]^{\dagger}=o(N^{-\gamma})
\] 
for all $n$. This yields in the Jordan basis
\begin{equation}\label{eBalanceCorJordan}
\sum_{m,k}[\mathcal{J}^{(n)}_{\jor}]_{\mu m}\hat{C}^{(n)}_{mk}[[\mathcal{J}^{(n)}_{\jor}]^{*}]_{k\nu}= o(N^{-\gamma})
\end{equation}
for all $\mu,\nu$. 

By definition of $P_{\max}$, for at least one Fourier mode, $n$, the matrix $\boldsymbol{\mathcal{J}}^{(n)}_{\jor}$ has at least one block whose real part is a shift matrix of degree $P_{\max}$. In general, there can be several such Jordan blocks. For example, in the network depicted in Fig.~\ref{fig:4pop}b, for which $P_{\max}=2$, there are two Jordan blocks with $P=2$.   

We first assume that all blocks which are a shift matrix of size $P_{max}$ are real. Since for such blocks in $\boldsymbol{\mathcal{J}}_{\jor} \boldsymbol{\hat{C}}\boldsymbol{\mathcal{J}}^{*}_{\jor}$ scale the same with $K$, it is sufficient to consider the case where there is only one such block.
We denote it by $\boldsymbol{S}$ and by $\boldsymbol{\hat{C}}_{\max}$ the corresponding block in $\boldsymbol{\hat{C}}$. Equation \eqref{BalanceCorrStrong} then yields
\begin{equation}\label{eBalanceCorJordan2}
\boldsymbol{S}\boldsymbol{\hat{C}}_{\max}\boldsymbol{S}=o(N^{-\gamma})
\end{equation}
For instance, for $P_{\max}=2$, we have (see Eq.~\eqref{degen})
\[
\boldsymbol{\hat{C}}_{\max}=\left( \begin{array}{cc}
\mathcal{O}(N^{\gamma/2-1}) & \mathcal{O}(N^{\gamma-1})  \\
0 & \mathcal{O}(N^{\gamma/2-1}) \\
\end{array} \right)
\]
and thus
\[
\boldsymbol{S}\boldsymbol{\hat{C}}_{\max}\boldsymbol{S}=0
\]
Therefore, for this block Eq.~\eqref{BalanceCorrStrong} is always satisfied. The latter equation, however, also applies to other Jordan blocks and Fourier modes. This implies that $\gamma < \gamma_{\max}=1$

For $P_{\max}=3$, we have
\[
\boldsymbol{\hat{C}}_{\max}=\left( \begin{array}{ccc}
\mathcal{O}(N^{\gamma-1}) & \mathcal{O}(N^{3\gamma/2-1}) & \mathcal{O}(N^{2\gamma-1}) \\
\mathcal{O}(N^{\gamma/2-1}) & \mathcal{O}(N^{\gamma-1}) & \mathcal{O}(N^{3\gamma/2-1}) \\
0 & \mathcal{O}(N^{\gamma/2-1}) & \mathcal{O}(N^{\gamma-1}) \\
\end{array} \right)
\]
and
\[
\boldsymbol{S}\boldsymbol{\hat{C}}_{\max}\boldsymbol{S}=\left( \begin{array}{ccc}
0 & \mathcal{O}(N^{\gamma/2-1}) & \mathcal{O}(N^{\gamma-1}) \\
0 & 0 & \mathcal{O}(N^{\gamma/2-1}) \\
0 & 0 & 0 \\
\end{array} \right)
\]
Thus, Eq.~\eqref{BalanceCorrStrong} is satisfied only if $\gamma < \gamma_{\max}=1/2$.

In general, for a $P_{\max}\times P_{\max}$ shift matrix $\boldsymbol{S}\boldsymbol{\hat{C}}_{\max}\boldsymbol{S}$ is $\mathcal{O}(N^{\gamma(P_{\max}-2) -1})$. This implies that $\gamma<\gamma_{\max}$ with
\begin{equation}
\gamma_{\max}=\frac{1}{P_{\max}-1}
\end{equation}.

\medskip

Let us now consider networks in which there is at least one pair of complex conjugate Jordan blocks whose real parts are a shift matrix of size $P_{\max}$. An example, of such a network is depicted in Fig.~\ref{fig:4pop}d. The first Fourier mode of the population averaged connectivity matrix in this 
example is
\begin{equation}\label{e4popExample}
\boldsymbol{J}^{(1)}=\left( \begin{array}{cccc}
0 & -a & c & 0 \\
b & 0 & 0 & 0 \\
0 & 0 & 0 & -a \\
0 & 0 & b & 0 \\
\end{array} \right),
\end{equation}
with $a,b,c$ real and positive. For this network
\begin{equation}\label{eComplexJordan}
\boldsymbol{\mathcal{J}}_{\jor}=\left( \begin{array}{cccc}
i\omega & 1 & 0 & 0 \\
0 & i\omega & 0 & 0 \\
0 & 0 & -i\omega & 1 \\
0 & 0 & 0 & -i\omega \\
\end{array} \right)
\end{equation}
with $\omega=\sqrt{ab}$. 

These complex conjugate blocks can in general be written as $\pm i \omega \boldsymbol{I} + \boldsymbol{S}$, where $\boldsymbol{I}$ is the identity matrix of size $P_{\max}$. Thus 
\begin{eqnarray}
\mathcal{\boldsymbol{J}}_{\jor}^{\max} \boldsymbol{\hat{C}}_{\max} \mathcal{[\boldsymbol{J}}_{\jor}^{\max}]^* &=&  \omega^{2} \boldsymbol{\hat{C}}_{\max} \pm i \omega [\boldsymbol{\hat{C}}_{\max} \boldsymbol{S} -\boldsymbol{S} \boldsymbol{\hat{C}}_{\max}]
\nonumber \\&& +\boldsymbol{S} \boldsymbol{\hat{C}}_{\max} \boldsymbol{S}
\end{eqnarray}

As shown above, $\boldsymbol{\hat{C}}_{\max}=\mathcal{O}(N^{\gamma (P_{\max}-1)-1})$ and $\boldsymbol{S} \boldsymbol{\hat{C}}_{\max} \boldsymbol{S}=\mathcal{O}(N^{\gamma (P_{\max}-2)-1})$.
It is straightforward to also show that $\boldsymbol{\hat{C}}_{\max} \boldsymbol{S} -\boldsymbol{S} \boldsymbol{\hat{C}}_{\max}= \mathcal{O}(N^{\gamma (P_{\max}-3/2)-1})$. Therefore $\mathcal{\boldsymbol{J}}_{\jor}^{\max} \boldsymbol{\hat{C}}_{\max} \mathcal{[\boldsymbol{J}}_{\jor}^{\max}]^*=\mathcal{O}(N^{\gamma (P_{\max}-1)-1})$. Equation \eqref{BalanceCorrStrong} is then satisfied only if $\gamma < \gamma_{\max}$, with
\begin{equation}\label{gammacond2}
\gamma_{\max} =\frac{1}{P_{\max}}
\end{equation}

According to Theorem 2, if  $\boldmath{\mathcal{J}}^{(n)}_{\jor}$ has a block whose real part is a shift matrix for  at least  one mode $n$, $\boldsymbol{C} =\mathcal{O}(N^{-\alpha})$ where $\alpha=1-\gamma(P_{\max} -1)$. If $\gamma < \gamma_{\max}$, correlations in the activity will decrease more slowly than $1/N$, when $N$ is increased.
If $\gamma > \gamma_{\max}$ correlations will increase with $N$ and the network will not operate in the balanced regime. Finally, if $\gamma=\gamma_{\max}$, our theory will give $\mathcal{O}(1)$ correlations in the input which is inconsistent with the Ansatz in Eq.~\eqref{BalanceCorrStrong}. In this case substantial corrections to the 
Eq.~\eqref{CorrEqMat} should be taken into account. A different approach, similar to the one in \cite{renart2010asynchronous} must be adopted to self-consistently determine these correlations.

\section{Discussion}\label{SDiscussion}

\subsection{Main results}

We developed a theory for the emergence of correlations in  strongly recurrent networks of binary neurons. Each neuron receives on average $K$ inputs from each of $D$ populations, with probabilities which are spatially modulated. The synaptic strengths scale as $1/\sqrt{K}$ and the  network operates in the balanced regime \cite{van1996chaos}. For simplicity we considered networks with a one dimensional ring architecture with  connection probabilities solely dependent on distance and on the nature (excitatory or inhibitory) of the pre- and postsynaptic populations. 

We present a balanced correlation equation, which together with the balanced rate equation, define the balanced regime and insure that mean inputs to the neurons and their fluctuations are both $\mathcal{O}(1)$. We derive a set of equations that determine the equilibrium values of the quenched averaged correlations and we show that the solution of these equations is stable provided that the solution of the balanced rate equations is stable.

Key results of our work are two scaling correlation theorems. The first shows that generically, all the Fourier modes of the quenched average correlations are small when $N$ and $K$ are large. They are of $\mathcal{O}(1/N)$, and independent of $K$ to leading order. This is true in the large $N$ limit even if we take $K=pN$, provided that $p$ is not too large. However, the second theorem states that there are recurrent network architectures in which some of the Fourier modes of the quenched averaged correlations increase with $K$. These architectures are characterized by an explicit, or a hidden, feedforward structure in those modes. This structure is revealed by the Jordan form of the interaction matrix averaged over realizations. If this Jordan form contains a block whose real part is a shift matrix  of size $P>1$, the corresponding mode in the correlation increases at least as $\mathcal{O}(\frac{K^{P-1}}{N})$. Importantly, in these cases the network still operates in the balanced regime provided that $K\lesssim N^{\frac{1}{P-1}}$ (or $K\lesssim N^{\frac{1}{P}}$, see Section \ref{SlargeK}). Corrections to the theory become important when $K$ approaches $N^{\frac{1}{P-1}}$ (or  $N^{\frac{1}{P}})$.

\subsection{Generality of the results}

For simplicity, we assumed that synaptic weights depend only on the identities of the populations to which pre and postsynaptic neurons belong. The results, however, will not change if the weights are heterogeneous with distributions that depend on the pre and postynaptic populations, as long as the mean and the variance of all these distributions are finite.

For notational simplicity we assumed that all populations have the same number of neurons, $N$, and each population receives, on average, inputs from $K$ neurons in every population. The theory can be easily extended  to networks in which population $\alpha$ has $N_\alpha=\nu_\alpha N$ neurons and the average number of connections from population $\beta$ to population $\alpha$ is $K_{\alpha \beta}=\kappa_{\alpha \beta} K$. This will not affect the scaling of the correlations with $N$ and $K$. Prefactors, however, will be different. For instance, assuming four times fewer inhibitory than excitatory neurons in the two-population networks of subsection \ref{SS2popGeneral}, without changing the number of connections per neuron, the correlations of the inhibitory neurons will increase by a factor of 4.

We focused on networks with a one-dimensional ring architecture and connection probabilities which are solely distance dependent. This greatly simplifies the problem, because when averaged over realizations, correlations depend solely on distance. Furthermore, because of the linearity of the self-consistent equation for the correlations, the different Fourier modes decouple, allowing us to analyze each mode separately. However, our analytical approach does not require rotation invariance. It can be extended to any network architecture for which the Jordan normal form of the interaction matrix, averaged over realizations, can be established.

\subsection{Robustness and self-consistency of the results}
The theory presented here makes the Anzatz that correlations are sufficiently small so that the dynamics of the crosscorrelations can be linearized (Appendix 
\ref{ApCorBinary}). If this Anzatz is correct, the theory is self-consistent. When the theory predicts correlations which are $\mathcal{O}(1)$, non-linear terms contribute and the correlations start to deviate from the theoretical value (see for example simulation results for $N=40000$ in Fig.~\ref{fig:BinaryEI}b). Nevertheless, the order of the correlations is still correctly predicted.

In the large $N,K$ limit, only networks with feedforward structures  -hidden or explicit- can exhibit correlations that increase with the average number of inputs.  However, when $K$ and $N$ are only moderately large, as is the case in biological systems, a strict tuning of the architecture is not necessary. This is because there is a crossover between the regimes of strong and weak correlations as $K$ is increased (see Fig.~\ref{fig:BinaryEI-II}c). As shown in Appendix \ref{ApProof}\ref{ApSs1}, the value of $K$ for which this crossover occurs depends on the eigenvalues of the interaction matrix.

\subsection{Relation to previous works}

Non-interacting neurons can exhibit correlations if they share feedforward inputs \cite{shea2008correlation,de2007correlation}. For instance, Litvak et al. \cite{litvak2003transmission} investigated a chain of layers of integrate-and-fire neurons lacking any recurrent interactions and coupled only feedforwardly. In their model, each neuron in a layer receives inputs from the same number of excitatory and inhibitory neurons in the previous layer in such a way that their temporal averages {\it exactly} balance. They found a build up of correlations along the chain. This is because the correlations induced by shared feedforward inputs are not suppressed during the activity propagation  since the network lacks any recurrent interactions and is thus purely feedforward.

Cortes and van Vreeswijk \cite{cortes2015pulvinar} studied a chain of strongly recurrent unstructured  E-I subnetworks coupled through  excitatory unstructured feedforward projections. They found a gradual build up of correlations along the chain. These correlations, however, decrease if the connectivity, $K$, and the sub-networks size, $N$, increase together. This result is in agreement with our theory which predicts that the correlations are $\mathcal{O}(1/N)$ through the whole chain.

In \cite{ginzburg1994theory} Ginzburg and Sompolinsky considered networks of binary neurons with finite temperature Glauber dynamics, unstructured, dense ($K=pN$) connectivity and weak interactions, i.e., of the order of $\mathcal{O}(1/K)$ and not $\mathcal{O}(1/\sqrt{K})$ as in our work. Mean Field theory shows that in these networks correlations are $\mathcal{O}(1/N)$ with a prefactor which diverges in the zero temperature limit. This is in contrast to what happens in the strongly recurrent unstructured networks we considered here where correlations also scale as $\mathcal{O}(1/N)$ but with a prefactor, which is finite despite the fact that we assumed zero temperature Glauber dynamics. This is because in strongly recurrent networks, intrinsic noise emerges from the deterministic dynamics of the network.

They also demonstrated that in their model the correlations amplify up to $\mathcal{O}(1)$ at Hopf bifurcation onsets \cite{ginzburg1994theory}. At such onsets the dynamics exhibit critical slowing down and thus this amplification is accompanied by a divergence of the decorrelation times. Our work demonstrates a different amplification mechanism: it occurs at a point where the Jordan form of the interaction matrix, $\boldsymbol{\mathcal{\bar{J}}}$, contains a block which is a shift matrix. Since there is no critical slowing down at such a point, the decorrelation times are finite and on the order of the update time constant (data not shown). Our theory may thus account for substantial correlations with short time scales, as frequently observed in the brain \cite{smith2008spatial,smith2013laminar,renart2010asynchronous}. 

Renart et al. \cite{renart2010asynchronous} and Helias et al. \cite{helias2014correlation} investigated strongly recurrent unstructured networks with one excitatory and one inhibitory population of binary neurons which are densely connected, i.e with $K \propto N$.  They found that in these networks mean pair correlations  were $\mathcal{O}(1/N)$. As they only considered dense connectivity they could not, however, disentangle the dependence on $N$ and $K$. Here, we considere different relations between $K$ and $N$ and show that in unstructured networks the correlations are $\mathcal{O}(1/N)$, and in practice do not depend on $K$. This last result is remarkable since one would expect correlations to increase with the degree of connectivity. This is not the case: the balance of excitation and inhibition prevents that to occur in unstructured networks.

On the other hand, and somewhat surprisingly, we found that even if the fraction of common inputs shared by neurons is very small, a build up of correlations can still occur for some network architectures. For example, in a network of four populations with a feedforward structure, correlations would be of $\mathcal{O}(K^3/N)$, and thus, to satisfy the balanced correlation equation, the scaling of $K$ with the network size can be at most $K=\mathcal{O}(N^{1/3})$. However, with this architecture and scaling, the probability of two neurons to share their inputs is  $\mathcal{O}(N^{-4/3})$ (Eq.~\eqref{common}) and the number of inputs shared by two neurons is therefore $\mathcal{O}(N^{-1/3})$. Thus, although the number of shared inputs goes to zero in the large N limit, correlations get amplified up to $\mathcal{O}(1)$  thanks to the feedforward architecture.

Rosenbaum et al. \cite{rosenbaum2017spatial} have recently investigated how feedforward excitation can drive correlations in spatially structured E-I networks operating in the balanced regime. The specific architecture they considered is reminiscent of the particular example presented in Fig.~\ref{fig:4pop}a. In their study the fluctuations which drove the correlated activity were those in the feedforward inputs and the contribution to the correlations of the fluctuations generated by the recurrent dynamics was neglected. In contrast, our work focuses on the role of the {\it recurrent} dynamics in the emergence of correlations.

We recently studied the emergence of correlations in a  network consisting of two strongly recurrent E-I sub-networks, the first projecting to the second with topographically organized feedforward connections \cite{darshan2017canonical}. The architecture in that work is also reminiscent of the example presented in Fig.~\ref{fig:4pop}a. We showed that while in the first subnetwork correlations were weak, $\mathcal{O}(1/N)$, in the second subnetwork the activity was self-organized in such a way that correlations in macroscopic sub-populations of excitatory neurons were finite and did not depend on $N$ and $K$ when the latter were sufficiently large. This does not contradict our theory. In the architecture considered in \cite{darshan2017canonical} subsets of neurons in the first subnetwork are projecting to a macroscopic fraction of neurons in the second subnetwork. This organization generates correlations between elements of the connectivity matrix, unlike in the models considered here. Generalizing our theory to such cases is possible but beyond the scope of this paper.

\subsection{Directions for future work}

The theorems of Section \ref{Stheorem}, tell us how the Fourier modes of the averaged correlations scale for large $N$ and $K$. In the examples considered in Sections \ref{S2networks}, \ref{AppliTh} and \ref{SlargeK}, we focused on connectivities whose Fourier expansion involve only two modes. In these cases, the scaling of the spatial correlations can be immediately deduced from that of the Fourier modes. This will also be the case for interactions described by a finite number of Fourier modes. However, if the interactions are described by an infinite number of  modes, inferring the scaling of the correlations from those of their modes can be more complicated. For instance, if one takes the large $N$ limit with $K \propto N^{\gamma}$, the convergence of the Fourier series may not be uniform in $N$. In that case the scaling with $N$ of the spatial correlations may be highly non-trivial. We will address this issue in an upcoming paper. 

The present paper focuses on locally averaged two-point correlation functions. Recent progress in experimental techniques will create large data sets of neuronal activities from which distributions of pairwise correlations can be extracted. The power of theoretical approaches to interpret such data will be greatly enhanced if they provide not only locally averaged correlations but also higher order statistics of their distributions. Thus it would be interesting to extend our approach to estimate the scaling of higher order moments of correlations in binary networks.

Several previous studies investigated  EI networks ({\it e.g}. \cite{helias2014correlation,brunel2000dynamics}) in which inhibition and excitation were unstructured and their strengths were only a function of the presynaptic neurons, i.e., $J_{EE}=J_{IE}$ and $J_{II}=J_{EI}$. Other studies assumed unstructured connectivity  with  $J_{\alpha E}=-J_{\alpha I}$, $\alpha\in\{E,I\}$ ({\it e.g.} \cite{rajan2006eigenvalue}). In both cases the interaction parameters are on the edge of the region where the network evolves towards the balanced state (see Eq.~\eqref{eBalanceConditions}). The network dynamics may be qualitatively different on the edge of this region than inside it. It is thus not clear that the scaling theorems presented here apply to these tuned cases. A further investigation of the correlation structure in such networks is a subject for future research.

Do the conclusions derived here for networks of binary neurons hold for networks with more realistic single neuron dynamics? To approach this question we performed extensive numerical simulations of strongly recurrent networks consisting of one inhibitory and one excitatory population of leaky integrate-and-fire (LIF) neurons. The detailed analysis of these simulations will be presented elsewhere. In brief, we found in our simulations, that in these networks the averaged pairwise correlations scale with $K$ and $N$  in manner that is consistent with the theory presented here for binary networks. It would be very interesting to extend our analytical approach to these type of networks.

To conclude, van Vreeswijk and Sompolinsky \cite{van1996chaos,vreeswijk1998chaotic} investigated strongly recurrent networks of binary neurons with unstructured sparse connectivity. They showed how these network dynamics evolve into a balanced state. Due to the sparseness of the connectivity, correlations are negligible in these networks. Subsequent studies \cite{renart2010asynchronous,helias2014correlation} extended these results to unstructured networks with dense connectivity, and found that here too the network dynamics evolve to a balance state in which reverberations keep the correlations very small. In contrast, as shown here, in networks with structured connectivity correlations can be large. The theory presented here gives the conditions on the network architecture to evolve into a balanced state with strong correlations and show how they depend on the network size and number of connections.

\section*{acknowledgments}
We thank Gianluigi Mongillo and German Mato for fruitful discussions. Work conducted in the framework of the France Israel Laboratory of Neuroscience (FILN). Grants: ANR/CRCNS-BASCO, ANR-BALWM, ANR-BALAV1, France-Israel High Council for Science and Technology, LIA-FILN (CNRS), IRN-FICNC (CNRS).

\bibliography{SI}

\begin{appendix}

\section{Correlations in binary networks}\label{ApCorBinary}

Here we calculate the equilibrium value and the stability of the quenched average correlations.

We define, for $(j,\beta)\neq (i,\alpha)$, the out of equilibrium auto- and crosscorrelations, $c^{\alpha\beta}_{ij}(t,\tau)$ as 
\begin{eqnarray*}
c^{\alpha\beta}_{ij}(t,\tau) &\equiv&
\avIn{\delta S_i^\alpha(t)\delta S_j^\beta(t+\tau) }\\
a^{\alpha}_{i}(t,\tau) &\equiv&
\avIn{\delta S_i^\alpha(t)\delta S_i^\alpha(t+\tau) }
\end{eqnarray*}
where $\avIn{\cdot}$ denotes averaging over many initial conditions choosen with a probability measure that, for simplicity, we choose such that $\avIn{S_i^\alpha(0)}=\avT{S_i^\alpha(t)}$\@. It is also convenient for the notation to define
$c^{\alpha\alpha}_{ii}(t,\tau)=0$. 
In this paper we focus on the equal 
time correlations, $c^{\alpha\beta}_{ij}(t)\equiv c^{\alpha\beta}_{ij}(t,0)$. 

For networks of binary neurons the dynamics of the equal-time crosscorrelations is given by \cite{glauber1963time,ginzburg1994theory,renart2010asynchronous}
\begin{align*}
\frac{d c^{\alpha\beta}_{ij}(t)}{dt}= & -2c^{\alpha\beta}_{ij}(t)+\avIn{\delta \Theta[h_i^\alpha(t)-T]\delta S_j^\beta(t)}\\
	&+\avIn{\delta S_i^\alpha(t)\delta \Theta[h_j^\beta(t)-T]},
\end{align*}
where $\delta \Theta[h_i^\alpha(t)-T]\equiv
\Theta[h_i^\alpha(t)-T]-\avIn{\Theta[h_i^\alpha(t)-T]}$\@.

If we make the {\it Ansatz} that correlations are weak,
we can, to leading order, take
$\avIn{\delta \Theta[h_i^\alpha(t)-T]\delta S_j^\beta(t)}=g_i^\alpha \avIn{\delta h_i^\alpha(t)\delta S_j^\beta(t)} $,
where $g_i^\alpha$ is the gain of neuron $(i,\alpha)$,
$g_i^\alpha=\partial_h \avIn{\Theta[h_i^\alpha(t)-T]}$, 
which is, to leading order, independent of the 
correlations \cite{ginzburg1994theory,renart2010asynchronous,helias2014correlation}.

Thus,
\begin{eqnarray*}
\frac{d c^{\alpha\beta}_{ij}(t)}{dt} & = & 
-2c^{\alpha\beta}_{ij}(t)+g_i^\alpha\avIn{\delta h_i^\alpha(t)\delta S_j^\beta(t)} \\
   & & \mbox{}+g_j^\beta\avIn{\delta S_i^\alpha(t)\delta h_j^\beta(t)} \\
   & = & -2c^{\alpha\beta}_{ij}(t)+g_i^\alpha
\sum_{\gamma, k}J_{ik}^{\alpha\gamma}\avIn{\delta S_k^\gamma(t)\delta S_j^\beta(t)}+ \\
   & & \mbox{}+g_j^\beta \sum_{\gamma, k}J_{jk}^{\beta\gamma}\avIn{\delta S_k^\gamma(t)\delta S_i^\alpha(t)}.
\end{eqnarray*}
Using $\avIn{\delta S_i^\alpha(t)\delta S_j^\beta(t)}= 
c_{ij}^{\alpha\beta}(t)+\delta_{\alpha,\beta}\delta_{i,j}
a_i^\alpha(t)$ yields
\begin{eqnarray}
\frac{d c^{\alpha\beta}_{ij}(t)}{dt} & = & 
-2c^{\alpha\beta}_{ij}(t)+\sum_{\gamma, k} \left[
 \tilde{J}_{ik}^{\alpha\gamma} c_{kj}^{\gamma\beta}(t)
  + \tilde{J}_{jk}^{\beta\gamma} c_{ik}^{\alpha\gamma}(t) 
 \right] +\nonumber \\
   & & \mbox{ } + \tilde{J}^{\alpha\beta}_{ij} 
   a_j^\beta(t)+\tilde{J}^{\beta\alpha}_{ji}
   a_i^\alpha(t),
\label{GRH}
\end{eqnarray}
where $\tilde{J}^{\alpha \beta}_{ij}= g^\alpha_i J^{\alpha \beta}_{ij}$\@.
Note that, for weak correlations $\avIn{S_i^\alpha(t)}$ 
does not depend on the correlations so that, $\avIn{S_i^\alpha(t)}=\avT{S_i^\alpha(t)}$, 
for our choice of initial conditions ($\avIn{S_i^\alpha(0)}=\avT{S_i^\alpha(t)}$). 
Hence, the equal time autocorrelation $a_i^\alpha$ is 
independent of time and given by $a_i^\alpha=
\avT{S_i^\alpha(t)}-\avT{S_i^\alpha(t)}^2$\@.

We now average over the quenched disorder. Due to the 
rotational symmetry of the connection probabilities,
$P^{\alpha \beta}_{ij}$, $\avJ{a_i^\alpha}$ is a constant
$$\avJ{a_i^\alpha}=A_\alpha$$
whereas $\avJ{c^{\alpha\beta}_{ij}}$ and $\avJ{\tilde{J}^{\alpha \beta}_{ij}}$ are functions 
of the difference in the location of neurons $(i,\alpha)$ and $(j,\beta)$ 
$$\avJ{c^{\alpha\beta}_{ij}(t)}=C_{\alpha\beta}(\theta_i^\alpha-\theta_j^\beta,t)$$ 
$$\avJ{\tilde{J}^{\alpha \beta}_{ij}}=\frac{1}{N} \mathcal{J}_{\alpha\beta}(\theta_i^\alpha-\theta_j^\beta)$$ where $\mathcal{J}_{\alpha\beta}(\Delta)= \sqrt{K}g_\alpha J_{\alpha\beta}
f_{\alpha\beta}(\Delta)$. Here we have assumed that the correlations in the quenched disorder in the inputs to the neurons are small, such that, to leading order, the expected value of the gain does not depend on the neuronal position. We comment on this Ansatz in Appendix \ref{ApQuenched}.

Thus, for large $N$, the quenched average of 
Eq.~\eqref{GRH} yields
\begin{align}\label{corfour}
\frac{d}{dt}C_{\alpha\beta}(\Delta,t)= &
-2C_{\alpha\beta}(\Delta,t)+\sum_\gamma\int\!
\frac{d\Delta^\prime}{2\pi} \nonumber \\
&\mbox{}\Big[\mathcal{J}_{\alpha\gamma}(\Delta-\Delta^\prime)
C_{\gamma\beta}(\Delta^\prime,t)+\Big. \nonumber \\
&\mbox{}+\Big.\mathcal{J}_{\gamma\beta}(-\Delta^\prime)
C_{\alpha\gamma}(\Delta-\Delta^\prime,t)\Big]+ \nonumber \\
  & \mbox{}+\mathcal{J}_{\alpha\beta}(\Delta)
    \frac{A_\beta}{N}+
    \mathcal{J}_{\beta\alpha}(-\Delta)
    \frac{A_\alpha}{N}.
\end{align}
Here we have not taken into account that $c_{ii}^{\alpha\alpha}(t)=0$\@. It is easy to see, however, that in the cross-correlations this neglects $\mathcal{O}(1/N^2)$ corrections. We show below that these corrections are indeed negligeable in the large $N$ limit.

The $n^{th}$ Fourier mode of $C_{\alpha\beta}(\Delta,t)$, $C_{\alpha\beta}^{(n)}(t)$, satisfies
\begin{align*}
\frac{d}{dt}C_{\alpha\beta}^{(n)}(t)= &
-2C_{\alpha\beta}^{(n)}(t)\\
&\mbox{}+\sum_\gamma
\Big[\mathcal{J}_{\alpha\gamma}^{(n)}
C_{\gamma\beta}^{(n)}(t)\Big. + \Big.C_{\alpha\gamma}^{(n)}(t)[\mathcal{J}_{\gamma\beta}^{(n)}]^{ \dagger}\Big] \\
  & \mbox{}+\mathcal{J}_{\alpha\beta}^{(n)}
    \frac{A_\beta}{N}+
    \frac{A_\alpha}{N}
    [\mathcal{J}_{\alpha\beta}^{(n)}]^{\dagger},
\end{align*}
where $\mathcal{J}_{\alpha\beta}^{(n)}$ and 
$[\mathcal{J}_{\alpha\beta}^{(n)}]^{\dagger}$ are the 
$n^{th}$ Fourier mode of
$\mathcal{J}_{\alpha\beta}(\Delta)$ and 
$\mathcal{J}_{\alpha\beta}^\top(\Delta)$, respectively.

This can be written more compactly as
\begin{align}
\tau\frac{d}{dt}\boldsymbol{C}^{(n)}(t)= &
-2\boldsymbol{C}^{(n)}(t)+
\boldsymbol{\mathcal{J}}^{(n)}
\boldsymbol{C}^{(n)}(t)+
\boldsymbol{C}^{(n)}(t)
[\boldsymbol{\mathcal{J}}^{(n)}]^\dagger \nonumber \\ 
  & \mbox{}+\boldsymbol{\mathcal{J}}^{(n)}
    \frac{\boldsymbol{A}}{N}+
    \frac{\boldsymbol{A}}{N}
    [\boldsymbol{\mathcal{J}}^{(n)}]^\dagger.
\end{align}
where $\boldsymbol{X}$ denotes the $D\times D$ matrix,
$X_{\alpha\beta}$, and $\boldsymbol{A}$ is the $D\times D$ matrix 
$A_{\alpha\beta}\equiv\delta_{\alpha,\beta} A_\alpha$. The equilibrium values of the Fourier  components of equal-time correlation functions thus satisfy Eq.~\eqref{CorrEqMat}.

\section{Correlation theorems}\label{ApProof}
In Appendix A we derived $N^2D^2$ coupled linear differential 
equations that determine the evolution of equal-time 
cross-correlations for all neuronal pairs in the network. Averaging these equations over the quenched disorder and using the rotation invariance of the connection probabilities yields a set of $ND^2$ coupled equations for the quenched averaged correlations. In Fourier space these equations lead to $N$ independent sets of $D^2$ coupled equations for the correlations. Here we prove Theorems 1 and 2 (see Section~\ref{Stheorem}) which state how these correlations scale with $N$ and $K$. 

To leading order, $\boldsymbol{A}$ is independent of $K$ and $N$, but $\boldsymbol{\mathcal{J}}^{(n)}$  is proportional to $\sqrt{K}$\@. Accordingly, we define
$\boldsymbol{\bar{\mathcal{J}}}^{(n)}\equiv
\boldsymbol{\mathcal{J}}^{(n)}/\sqrt{K}$ and
$[\boldsymbol{\bar{\mathcal{J}}}^{(n)}]^\dagger \equiv
[\boldsymbol{\mathcal{J}}^{(n)}]^\dagger/\sqrt{K}$\@. 
Thus, we can rewrite the evolution equation of the $n^{th}$ mode 
of the correlations as
\begin{align}
\frac{d}{dt}\boldsymbol{C}^{(n)}(t) = 
&-2\boldsymbol{C}^{(n)}(t)
\nonumber \\
  &+\sqrt{K}\left[\boldsymbol{\bar{\mathcal{J}}}^{(n)}
  \boldsymbol{C}^{(n)}(t)+
  \boldsymbol{C}^{(n)}(t)
   [\boldsymbol{\bar{\mathcal{J}}}^{(n)}]^\dagger \right]
 \nonumber \\
   &+ \frac{\sqrt{K}}{N}\left[
\boldsymbol{\bar{\mathcal{J}}}^{(n)}
    \boldsymbol{A}(t)+
    \boldsymbol{A}(t)
    [\boldsymbol{\bar{\mathcal{J}}}^{(n)}]^\dagger \right].
\label{scalecor}
\end{align}
The $D\times D$ matrix, $\boldsymbol{\bar{\mathcal{J}}}^{(n)}$ can be written as
\[
\boldsymbol{\bar{\mathcal{J}}}^{(n)}=\boldsymbol{U}^{(n)}
\boldsymbol{\mathcal{J}}^{(n)}_{\jor}[\boldsymbol{U}^{(n)}]^{-1}.
\]
where $\boldsymbol{\mathcal{J}}^{(n)}_{\jor}$ is the Jordan normal form 
of $\boldsymbol{\bar{\mathcal{J}}}^{(n)}$ and $[\boldsymbol{U}^{(n)}]^{-1}$ is the transformation matrix to the Jordan basis. 

The matrix $\boldsymbol{\mathcal{J}}^{(n)}_{\jor}$ can be written as
\[
[\boldsymbol{\mathcal{J}}^{(n)}_{\jor}]_{\mu\nu}
=\lambda^{(n)}_\mu\delta_{\mu,\nu}+\epsilon^{(n)}_\mu\delta_{\mu,\nu-1}.
\]
where $\lambda^{(n)}_\mu$ are the eigenvalues of 
$\boldsymbol{\bar{\mathcal{J}}}^{(n)}$ and $\epsilon^{(n)}_\mu=1$ inside a Jordan block and is 0 otherwise
(for clarity, in the Jordan basis we use the subscripts $\mu$ 
and $\nu$, rather than $\alpha$ and $\beta$ that we use in the original basis).

Importantly, the Jordan form of a matrix and of its Hermitian conjugate are complex conjugate. We thus can write
\[
[\boldsymbol{\bar{\mathcal{J}}}^{(n)}]^\dagger=\boldsymbol{V}^{(n)}
[\boldsymbol{\mathcal{J}}^{(n)}_{\jor}]^{*}[\boldsymbol{V}^{(n)}]^{-1},
\]

For notational convenience we will suppress the superscript
$(n)$ in the rest of this Appendix.
Defining $\boldsymbol{\hat{C}}$ as 
\[
\boldsymbol{\hat{C}}=\boldsymbol{U}^{-1}\boldsymbol{C}
\boldsymbol{V}
\]
and inserting into Eq.~\eqref{scalecor} yields
\begin{align}
\frac{d}{dt} \hat{C}_{\mu\nu}(t)= & 
-\Lambda_{\mu\nu}\hat{C}_{\mu\nu}(t)+ \nonumber \\
 &\mbox{}\sqrt{K}\left[
\epsilon_\mu\hat{C}_{\mu+1,\nu}(t)+
\epsilon_{\nu-1}\hat{C}_{\mu,\nu-1}(t)
\right] \nonumber \\
 &\mbox{}+\frac{\sqrt{K}}{N}\left[
(\lambda_\mu+\lambda_\nu^*)\hat{A}_{\mu\nu}\right.+
\nonumber \\
 &\mbox{}+\left.\epsilon_\mu\hat{A}_{\mu+1,\nu}+
\epsilon_{\nu-1}\hat{A}_{\mu,\nu-1}\right],
\label{cor-jor}
\end{align}
where we have defined $\Lambda_{\mu\nu}=2-\sqrt{K}(\lambda_\mu+\lambda_\nu^*)$ and
\[
\boldsymbol{\hat{A}}=\boldsymbol{U}^{-1}
\boldsymbol{A}\boldsymbol{V}.
\]
We now assume that the connectivity is such that the
system is stable to perturbations of the locally averaged 
rates. This implies that the real part of all eigenvalues 
of $\boldsymbol{\bar{\mathcal{J}}}$ are less than $1/\sqrt{K}$. The real part of $\Lambda_{\mu\nu}$ is therefore positive and thus $\boldsymbol{\hat{C}}(t)$ converges to an equilibrium 
value, $\boldsymbol{\hat{C}}^\infty$.

To see this, first consider $\hat{C}_{D,1}(t)$ which
satisfies
\[
\frac{d}{dt}\hat{C}_{D,1}(t)=
-\Lambda_{D,1}\hat{C}_{D,1}(t)+
\frac{\sqrt{K}}{N}(\lambda_D+\lambda_1^*)\hat{A}_{D,1}.
\]
Therefore $\hat{C}_{D,1}(t)$, converges to its equilibrium value. 

The evolution equations of $\hat{C}_{D,2}$ can be written as
$$ \frac{d}{dt} \hat{C}_{D,2}(t)= -\Lambda_{D,2}\hat{C}_{D,2}(t)+M_{D,2}\hat{C}_{D,2}(t) $$
where $M_{D,2}$ depends on $\hat{C}_{D,1}(t)$, $\hat{A}_{D,1}$ and $\hat{A}_{D,2}$. Since $\hat{C}_{D,1}(t)$ converges to its equilibrium value, $M_{D,2}$ converges to a constant.
Because $\mathrm{Re}(\Lambda_{D,2})>0$, $\hat{C}_{D,2}$ also converges to its equilibrium value, $\boldsymbol{\hat{C}}_{D,1}^\infty$. A similar argument shows that likewise $\hat{C}_{D-1,1}$  converges to $\hat{C}_{D-1,1}^\infty$. One then sees by recursion that the whole matrix $\boldsymbol{\hat{C}}$ converges to its equilibrium value, $\boldsymbol{\hat{C}}^\infty$. Hence,  Eq.~\eqref{CorrEqMat} determines the {\it stable} equilibrium values of the correlations.

From here we only consider the correlations at equilibrium and, for notational simplicity, we drop the superscript $\infty$.  

In the Jordan basis the equilibrium values of the correlations satisfy 
\begin{eqnarray}\label{eqcinf} 
\Lambda_{\mu\nu}\hat{C}_{\mu\nu}&=&\sqrt{K}\left[
\epsilon_\mu\hat{C}_{\mu+1,\nu}+
\epsilon_{\nu-1}\hat{C}_{\mu,\nu-1}
\right] + \nonumber \\
 & & \mbox{}+\frac{\sqrt{K}}{N}\left[
(\lambda_\mu+\lambda_\nu^*)\hat{A}_{\mu\nu}+\right.
\nonumber \\
 & & \mbox{}\left. +\epsilon_\mu\hat{A}_{\mu+1,\nu}+
\epsilon_{\nu-1}\hat{A}_{\mu,\nu-1}\right],
\label{cor-jor-eq}
\end{eqnarray}

Let us now consider the case where
$\boldsymbol{\bar{\mathcal{J}}}$ is diagonizable, so 
that $\epsilon_{\mu}=0$ for all $\mu$. In this case $\hat{C}_{\mu\nu}=0$ for $\mu \neq \nu$ and
\[
\hat{C}_{\mu\mu}=\frac{\sqrt{K}(\lambda_{\mu}+\lambda_{\mu}^*)}{N[2-\sqrt{K}(\lambda_{\mu}+\lambda_{\mu}^*)]}
\hat{A}_{\mu\mu},
\]
Thus
\begin{equation}
\lim_{K\rightarrow \infty} \left(\lim_{N\rightarrow \infty} N \hat{C}_{\mu\mu} \right) = -\hat{A}_{\mu\mu}
\end{equation}
unless $|\lambda_{\mu}+\lambda_{\nu}^*|$ is ${\scriptstyle\mathcal{O}}(1/\sqrt{K})$, in which case
\begin{equation}
\lim_{K\rightarrow \infty} \left(\lim_{N\rightarrow \infty} N \hat{C}_{\mu\nu}\right) = 0
\end{equation}
In the first situation $\hat{C}_{\mu\mu}=\mathcal{O}(1/N)$. In the second
situation  $\hat{C}_{\mu\mu}={\scriptstyle\mathcal{O}}(1/N)$\@. 
 
Let us now assume that $\boldsymbol{\bar{\mathcal{J}}}$ is not diagonizable. Then the $D\times D$ Jordan form of $\boldsymbol{\bar{\mathcal{J}}}$ consists of $B$ Jordan blocks  ($1 \leq B < D$) that we denote by $[\boldsymbol{\bar{\mathcal{J}}}_{\jor}]^i,i=1,...,B$. The size of the $i$th block will be denoted by $s(i)\times s(i)$. Without loss of generality, we can assume that the blocks are ordered in increasing size. 

The indices $\mu$ and $\nu$ of the elements of the Jordan block $i$, take values between
$l(i)$ and $h(i)$, with $l(i)=1+\sum_{j=1}^{i-1}s(j)$ and
$h(i)=\sum_{j=1}^i s(j)$\@. All the diagonal elements of this Jordan block are equal one of the eigenvalues of $\boldsymbol{\mathcal{J}}$, which we denote by $\hat{\lambda}_i$. The off-diagonal elements, $\epsilon_{\mu}$, are all equal to $1$ except for $\mu=h(i)$ ($i=1,..,B-1$) for which $\epsilon_{\mu}=0$\@.  

The matrix $\boldsymbol{\hat{C}}$ consists of $B^2$ sectors that we denote by 
$S_{ij}$. In the sector $S_{ij}$, $\mu \in  \{l(i), \ldots ,h(i)\}$ and $\nu \in \{l(j), \ldots ,h(j)\}$\@.

Let us consider Eq.~\eqref{eqcinf} for $\mu,\nu$ in sector $S_{ij}$. Since $\epsilon_{h(i)}$ and $\epsilon_{l(j)-1}$ are zero, the equation in this sector does not
depend on elements of $\boldsymbol{\hat{C}}$ outside of it. Thus, we can solve Eq.~\eqref{eqcinf} recursively to determine all the elements of $\boldsymbol{\hat{C}}$ in this sector. The recursion goes as follows. First, one solves for $\hat{C}_{h(i),l(j)}$
\begin{equation}
\hat{C}_{h(i),l(j)}= \frac{1}{N} \frac{\sqrt{K}\left(\hat{\lambda}_i+\hat{\lambda}_{j}^*\right)}{2-\sqrt{K}\left(\hat{\lambda}_i+\hat{\lambda}_{j}^* \right)} \hat{A}_{h(i)l(j)}
\end{equation}
We can the solve Eq.~\eqref{eqcinf} to get $\hat{C}_{\mu\nu}$ for $\mu,\nu =h(i)-1,l(j)$  and $\mu,\nu =h(i),l(j)+1$. This process can be repeated until all the elements of $\boldsymbol{\hat{C}}$ in the sector $S_{ij}$ are determined. 

\subsection{$\hat{A}_{h(i),l(j)}$ are all non-zero}\label{ApSs1}

A similar recursion can be performed to estimate the order of magnitude of all the elements of $\boldsymbol{\hat{C}}$. 
First we obtain
\[
\hat{C}_{h(i),l(j)}=
\mathcal{O}\left(\frac{\hat{A}_{h(i),l(j)}}{N}\right)=
\mathcal{O}\left(\frac{1}{N}\right).
\]
For $\hat{\lambda}_i+\hat{\lambda}_j^*=\mathcal{O}(1)$ the recursion shows that we have
\begin{equation}\label{Gen}
\hat{C}_{\mu\nu}=\mathcal{O} \left(\frac{1}{N}\right)
\end{equation}
for all $\mu, \nu$ in sector $S_{ij}$.

If $\hat{\lambda}_i+\hat{\lambda}_j^*=0$, $\hat{C}_{h(i),l(j)}=0$ and by recursion all 
the other elements in sector $S_{ij}$ are
\begin{equation}\label{degen}
\hat{C}_{\mu\nu}=\mathcal{O}\left(K^{\ell/2}/N\right),
\end{equation}
where $\ell=\nu-\mu+h(i)-l(j)$. Thus, in this case
 $\hat{C}_{l(i),h(j)}$ is the largest entry in the sector. It satisfies
 \begin{equation}\label{degenLargest}
 \hat{C}_{l(i),h(j)}=\mathcal{O}\left(K^{(s(i)+s(j)-2)/2}/N\right)
 \end{equation}
 
The scaling of the elements in sector $S_{ij}$ thus depends $\hat{\lambda}_i+\hat{\lambda}^*_j$. The stability of the dynamics imposes that $\mathcal{R}e(\hat{\lambda}_i)<0$ for all blocks (or positive but at most $\mathcal{O}(1/\sqrt{K})$, see next subsection) and thus the condition $\hat{\lambda}_i+\hat{\lambda}^*_j=0$  implies that, for large $K$, the real part of $\hat{\lambda}_i$ and $\hat{\lambda}_j$ are zero. In other words, the real part of $[\boldsymbol{\mathcal{J}}_{\jor}]^i$ and $[\boldsymbol{\mathcal{J}}_{\jor}]^j$ are shift matrices of size $s(i) \times s(i)$ and $s(j) \times s(j)$, respectively. For the sector $S_{ii}$ it means that the real part of $[\boldsymbol{\mathcal{J}}_{\jor}]^i$ is a shift matrix of size $s(i) \times s(i)$.

Proving Correlation Theorem 1 is now straightforward. If the real part of all the $B$ Jordan blocks of $\boldsymbol{\bar{\mathcal{J}}}$ are different from a shift matrix, one finds that $\hat{\lambda}_i+\hat{\lambda}^*_j=\mathcal{O}(1)\neq 0$ for all $i,j\in \{1,...,B\}$. According to Eq.~\eqref{Gen}, correlations are at most $\mathcal{O}(1/N)$ in all of the $B^2$ sectors of the matrix $\boldsymbol{\hat{C}}$. As a result, $\boldsymbol{C}^{(n)}=\mathcal{O}(1/N)$. On the other hand, if in each of the $B^2$ sectors of the matrix $\boldsymbol{C}$ correlations are at most $\mathcal{O}(1/N)$, this is also the case for the element of $\boldsymbol{\hat{C}}$ in all the sectors. In this case, Eq.~\eqref{Gen} implies that there is no Jordan block in $\boldsymbol{\bar{\mathcal{J}}}$ for which $\mathcal{R}e[\hat{\lambda}_i]=0$.

Restoring the index $n$ of the Fourier mode, one sees that if $\boldsymbol{\bar{\mathcal{J}}}^{(n)}$ has at least one Jordan block whose real part is a shift matrix and denoting by $P(n)$ the size of the largest shift, Eq.~\eqref{degenLargest} implies that $\hat{C}_{l(i),h(i)}=\mathcal{O}(K^{P(n)-1}/N)$, and thus also $\boldsymbol{\hat{C}}^{(n)}=\mathcal{O}(K^{P(n)-1}/N)$. This proves Correlation Theorem 2.

According to Eq.\eqref{degenLargest}, the scaling of the correlation is $\boldsymbol{\hat{C}}^{(n)}=\mathcal{O}(K^{D-1}/N)$ if and only if $P(n)=D$. When $\boldsymbol{\bar{\mathcal{J}}}^{(n)}$ is a real matrix (when the probability of connections are symmetric in $\Delta$), it means that the Jordan form of $\boldsymbol{\bar{\mathcal{J}}}^{(n)}$ is a shift matrix of size $D$. This is equivalent for saying that $\boldsymbol{\bar{\mathcal{J}}}^{(n)}$ is nilpotent of degree $D$. On the other hand, if $\boldsymbol{\bar{\mathcal{J}}}^{(n)}$ is a shift matrix of degree $D$, it has a Jordan form which is a shift matrix of degree $D$. According to Eq.\eqref{degenLargest}, the scaling of the correlation is then $\boldsymbol{\hat{C}}^{(n)}=\mathcal{O}(K^{D-1}/N)$. This proves the Corollary in section $\ref{Stheorem}$.

When $K=\mathcal{O}(1/|\hat{\lambda}_i+\hat{\lambda}_j^*|^2)$ there is a crossover in sector $S_{ij}$ of the matrix $\boldsymbol{\hat{C}}$ between weak correlations ($\mathcal{O}(1/N)$) and strong correlations ($\mathcal{O}(K^{(s(i)+s(j)-2)/2}/N)$).
To see this, we note that if $K=|\hat{\lambda}_i+\hat{\lambda}_j^*|^{-\frac{1}{\Gamma}}$, for $\Gamma \leq 1/2$
\[
\hat{C}_{\mu \nu}=\mathcal{O}\left(\frac{K^{\Gamma \ell}}{N}\right)
\]
whereas for $1/2 < \Gamma \leq 1$
\[
\hat{C}_{\mu \nu}=\mathcal{O} \left(\frac{K^{\ell/2}}{N}\right)
\]
for $\ell \neq 0$ and
\[
\hat{C}_{\mu \nu}=\mathcal{O}\left(\frac{K^{1/2-\Gamma}}{N}\right).
\]
for $\ell= 0$ 

\subsection{Some elements of $\boldsymbol{\hat{A}}$ are zero}\label{ApSs2}
So far, we have assumed that all the  elements in $\boldsymbol{\hat{A}}$ are non zero. The derivation can, however, be extended to include also situations where this is not the case as follows.  

Elements $\hat{A}_{\mu \nu}=0$ in a sector
where $\hat{\lambda}_i+\hat{\lambda}_j^*=\mathcal{O}(1)$ might result in
some elements of $\boldsymbol{\hat{C}}$ in that sector to be equal to 0
rather than $\mathcal{O}\left(\frac{1}{N}\right)$.\@
However, this will not affect the overall scaling of the correlations.

Elements $\hat{A}_{\mu \nu}=0$ in a sector $S_{ij}$ where $\hat{\lambda}_i+\hat{\lambda}_j^*=0$, do not change the scaling
in the matrix, provided that $\hat{A}_{h(i),l(j)} \neq 0$. If $\hat{A}_{h(i),l(j)} = 0$ the effect depends
on  $\hat{A}_{h(i)-1,l(j)}$ and $\hat{A}_{h(i),l(j)+1}$. If at least one of them is nonzero, the order of the  largest entry in the sector is decreased by $\sqrt{K}$. If both of them are zero, the order of this entry decreases at least by a factor of $K$. Reflecting on further element being zero, one reaches the following conclusion: Let $\{\mu_{ij},\nu_{ij}\}$ be the indices $\{\mu,\nu\}$ in $S_{ij}$ for which $\hat{A}_{\mu\nu}\neq 0$, which maximize $\mu-\nu$\@. Then, the maximal order of $\hat{C}_{\mu\nu}$ in the sector is $\mathcal{O}(K^{P_{ij}-1}/N)$, where
\begin{equation}\label{A0}
P_{ij}=1+(\mu_{ij}-l(i)+h(i)-\nu_{ij})/2.
\end{equation}

In conclusion: The highest order in
$\boldsymbol{\hat{C}}$ is $\mathcal{O}(K^{P-1}/N)$, where
$P$ is the maximum value of the $P_{ij}$  ($i,j\in\{1,\ldots,B\}$)
defined by: 1) $P_{ij}=1$ for sector $S_{ij}$ for which $\hat{\lambda}_i+\hat{\lambda}_j^*\neq 0$. 2) $P_{ij}=P$ for sector $S_{ij}$ in which $\hat{\lambda}_i+\hat{\lambda}_j^*=0$ and all $\hat{A}_{\mu \nu} \neq 0$\@. 3) $P_{ij}$ is given by Eq.~\eqref{A0} if in sector $S_{ij}$, $\hat{\lambda}_i+\hat{\lambda}_j^*=0$ but some $\hat{A}_{\mu \nu}$ are zero.

\section{Correlations of the quenched disorder}\label{ApQuenched}
Let us define the correlation of the quenched disorder in the outputs of the neurons as
\begin{equation}
\Gamma^{\alpha\beta}_{ij} = \avJ{\Delta S^{\alpha}_{i}\Delta S^{\beta}_{j}}
\end{equation}
for $(i,\alpha) \neq (j,\beta)$ and $\Delta S^{\alpha}_{i}$ defined in Eq.\eqref{Sseparate}.
A derivation similar to that in Appendix \ref{ApCorBinary} yields 
\begin{eqnarray}
2\Gamma^{\alpha\beta}_{ij}  & = & \sum_{\gamma, k} \left[ \mathcal{J}_{ik}^{\alpha\gamma} \Gamma_{kj}^{\gamma\beta}
  + \mathcal{J}_{jk}^{\beta\gamma} \Gamma_{ik}^{\alpha\gamma} 
 \right] +\nonumber \\
   & & \mbox{ } + \mathcal{J}^{\alpha\beta}_{ij} 
   q_\beta/N+\mathcal{J}^{\beta\alpha}_{ji}
   q_\alpha/N,
\label{GRH2}
\end{eqnarray}
where $q_\alpha=\avJ{\left( \Delta S^{\alpha}_{i}\right)^2}$. This equation can be solved using the same approach as in Appendices \ref{ApCorBinary}-\ref{ApProof}. This analysis shows that correlations of the quenched disorder and correlations of the temporal fluctuations are of the same order. Thus, when the temporal fluctuations are small, the Ansatz in Appendix \ref{ApCorBinary}, were we neglected the spatial fluctuations in the neuronal gain, is justified. If the correlations are too strong, the Ansatz may no longer be satisfied, but neither is the linearization assumed to derive equation \eqref{corfour}. 

\section{Self-consistent equations for the autocorrelation and the gain}\label{ApSCEqns}
We follow the notations from \cite{vreeswijk1998chaotic}:
\begin{equation}\label{SC1}
m_{\alpha}=\frac{1}{2} \erfc \left(\frac{T_{\alpha}-h_{\alpha}}{\sqrt{2(\sigma^2_{\alpha}+\sigma^2_{q\alpha})}}\right)
\end{equation}
\begin{equation}\label{SC2}
q_{\alpha}=\frac{1}{2}\int \!Dx\, \erfc^2 \left(\frac{T_{\alpha}-h_{\alpha}-\sigma_{q\alpha}x}{\sqrt{2\sigma_{\alpha}^2}}\right) 
\end{equation}
\begin{equation}\label{SC3}
A_{\alpha}=m_{\alpha}-q_{\alpha}
\end{equation}
where the variance of the input noise, $\sigma^2_{\alpha}=\avJ{\avT{   (h_i^{\alpha}-\avT{h_i^{\alpha}} )^2 }}$, is
\begin{widetext}
\begin{align}
\sigma^2_{\alpha}=\sum_{\beta} J^2_{\alpha\beta}(m_{\beta}-q_{\beta})
+K\sum_{\beta\beta'}\frac{J_{\alpha\beta}J_{\alpha\beta'}}{N^2}\sum_{j\neq j'}f_{\alpha\beta}(\theta_i^\alpha-\theta_j^\beta)f_{\alpha\beta'}(\theta_i^\alpha-\theta_{j'}^{\beta'})C_{\beta\beta'}(\theta_j^{\beta}-\theta_{j'}^{\beta'})
\end{align}
\end{widetext}
and the quenched disorder in the inputs, $\sigma^2_{q\alpha}=\avJ{ (\avT{h_i^{\alpha}}-\avJ{\avT{h_i^{\alpha}}} )^2 }$, is 
\\
\\
\begin{widetext}
\begin{equation}
\sigma^2_{q\alpha}=\sum_{\beta}J^2_{\alpha\beta}(q_{\beta}-p\int\frac{d\theta}{2\pi} f^2(\theta)m^2_{\beta})+K\sum_{\beta\beta'}J_{\alpha\beta}J_{\alpha\beta'}\frac{1}{N^2}\sum_{j\neq j'}f_{\alpha\beta}(\theta^\alpha_i-\theta^\beta_j)f_{\alpha\beta'}(\theta^\alpha_i-\theta^{\beta'}_{j'})  \avJ{ \Delta S^\beta_j\Delta S^{\beta'}_{j'} }
\end{equation}
\end{widetext}

Finally, the gain of the neurons is
\begin{equation}\label{SC5}
g_{\alpha}=\frac{1}{\sqrt{2\pi} \sigma_{T\alpha}}e^{\frac{-(T_{\alpha}-h_{\alpha})^2}{2(\sigma^2_{\alpha}+\sigma^2_{q\alpha})}}
\end{equation}
Equations \eqref{SC1}-\eqref{SC3} need to be solved self consistently. For simplicity, when we solved them we neglect $\mathcal{O}(p)$ terms (but see \cite{helias2014correlation}). 

\section{Cross-correlations in two-population networks}\label{Ap2pop}
For a two-population network Eq.~\eqref{CorrEqMat} yields
\begin{equation}
\boldsymbol{B}^{(n)} \left( \begin{array}{c}
C_{EE}^{(n)}\\
C_{EI}^{(n)}\\
C_{II}^{(n)}\\
 \end{array} \right)=\frac{1}{N}\boldsymbol{D}^{(n)} \left( \begin{array}{c}
A_E\\
A_I\\
  \end{array}  \right)
\end{equation}
with
\begin{equation*}
\boldsymbol{B}^{(n)}=\left( 
\begin{array}{ccc}
2(1-\mathcal{J}_{EE}^{(n)}) & -2\mathcal{J}_{EI}^{(n)} & 0 \\
-\mathcal{J}_{IE}^{(n)} &  2-(\mathcal{J}_{EE}^{(n)}+\mathcal{J}_{II}^{(n)}) & -\mathcal{J}_{EI}^{(n)}\\
0 & -2\mathcal{J}_{IE}^{(n)} & 2(1-\mathcal{J}_{II}^{(n)}) \\
 \end{array} \right)
\end{equation*}
and
\[
\boldsymbol{D}^{(n)}= \left( \begin{array}{cc}
2\mathcal{J}_{EE}^{(n)} & 0  \\
\mathcal{J}_{IE}^{(n)} & \mathcal{J}_{EI}^{(n)}\\
0 & 2\mathcal{J}_{II}^{(n)} \\
 \end{array} \right)
\]
Solving this equation one gets
\begin{widetext}
\begin{eqnarray}\label{Corfull2}
\nonumber
C^{(n)}_{EE}&=&\frac{1}{N}\frac{-2A_E\bar{\mathcal{J}}_{EE}^{(n)}\sqrt{K}+(-A_I(\bar{\mathcal{J}}_{EI}^{(n)})^2+A_E(\bar{\mathcal{J}}_{EE}^{(n)}T^{(n)}+\Delta^{(n)}+\bar{\mathcal{J}}_{EE}^{(n)}\bar{\mathcal{J}}_{II}^{(n)}) )K-A_E T^{(n)}\Delta^{(n)} K^{3/2}  }{-2+3T^{(n)}\sqrt{K}-((T^{(n)})^2+2\Delta^{(n)})K+T^{(n)}\Delta^{(n)} K^{3/2}}
\\ \nonumber
C^{(n)}_{EI}&=&-\frac{1}{N}\frac{(A_E\bar{\mathcal{J}}_{IE}^{(n)}+A_I\bar{\mathcal{J}}_{EI}^{(n)})\sqrt{K}-(A_I\bar{\mathcal{J}}_{EE}^{(n)}\bar{\mathcal{J}}_{EI}^{(n)}+A_E\bar{\mathcal{J}}_{II}^{(n)}\bar{\mathcal{J}}_{IE}^{(n)})K}{-2+3T^{(n)}\sqrt{K}-((T^{(n)})^2+2\Delta^{(n)})K+T^{(n)}\Delta^{(n)} K^{3/2}}
\\ \nonumber
C^{(n)}_{II}&=&\frac{1}{N}\frac{-2A_I\bar{\mathcal{J}}_{II}^{(n)}\sqrt{K}+(-A_E(\bar{\mathcal{J}}_{IE}^{(n)})^2+A_I(\bar{\mathcal{J}}_{II}^{(n)}T^{(n)}+\Delta^{(n)}+\bar{\mathcal{J}}_{EE}^{(n)}\bar{\mathcal{J}}_{II}^{(n)}) )K-A_I T^{(n)}\Delta^{(n)} K^{3/2}  }{-2+3T^{(n)}\sqrt{K}-((T^{(n)})^2+2\Delta^{(n)})K+T^{(n)}\Delta^{(n)} K^{3/2}}
\end{eqnarray}
\end{widetext}
where $T^{(n)}=Tr\boldsymbol{\bar{\mathcal{J}}}^{(n)}, \Delta^{(n)}=\det\boldsymbol{\bar{\mathcal{J}}}^{(n)}$. After some algebra, this equation can be rewritten as Eq.~\eqref{Corfull}.
\end{appendix}
\end{document}